# Drawing Lines in Psychological Space: What K-means Clustering Reveals in Simulated and Real Psychometric Data

**Pedro Henrique Ramos Pinto**[1]**, Maria Jullyanna Ferreira Marques**[2]**, Luiz Carlos Serramo Lopez**[1]

[1] Postgraduate Program in Cognitive and Behavioral Neuroscience (PPGNeC) - Federal University of Paraíba (UFPB) - João Pessoa, PB - Brazil

[2] Center for Health Sciences (CCS) - Federal University of Paraíba (UFPB) - João Pessoa, PB - Brazil

***Abstract.*** *K-means clustering is widely used in psychological and psychometric research to identify profiles, subgroups, and potential typologies, yet its classical formulation does not test whether such groups exist as latent psychological categories. Instead, K-means partitions multidimensional space into regions around centroids, favoring compact, approximately spherical clusters defined by geometric distance. In this paper, we examine this limitation through a sequence of controlled simulated datasets. We then extend the analysis to the SMARVUS dataset, a large international psychometric dataset comprising survey responses from university students across 35 countries, to evaluate whether similar geometric partitioning patterns emerge in empirical psychological data. By contrasting simulated and empirical data, this paper argues that K-means can produce stable and visually coherent clustering solutions even in continuous Gaussian latent spaces without true subgroup structure.*

## 1 Introduction

K-means is one of the most widely used unsupervised learning algorithms because it is intuitive, computationally efficient, and easy to interpret [Chong, 2021; Steinley, 2006]. This makes K-means especially attractive for person-centered analyses in psychology, psychiatry, education, and psychometrics, where researchers often aim to identify profiles, subgroups, or clinically relevant classifications [Gao et al., 2023]. For example, de la Fuente-Tomás et al. [2019] used K-means to develop a severity-based classification of patients with bipolar disorder, integrating cognition, functioning, physical health, and quality-of-life indicators into clinically interpretable profiles. Other studies have combined K-means with dimensionality reduction methods, generalized low-rank models, or Self-Organizing Maps (SOM) to explore psychopathology and behavioral heterogeneity [Grant et al., 2020; Croci et al., 2026].

Nevertheless, the interpretive status of K-means solutions remains a central methodological concern for nearly seven decades [Chong, 2021; Steinley, 2006]. Additionally, K-means does not test whether psychological types exist. It does not estimate latent categories, provide probabilistic class membership, or distinguish between discrete subgroups and continuous variation. Instead, it imposes a hard partition on a feature space by minimizing within-cluster distances around centroids [Raykov et al. 2016]. This is the basic assumption underlying the method: that useful partitions can be obtained by reducing within-cluster variance, not necessarily by demonstrating the existence of natural or latent categories.

This concern is particularly important because the accessibility and ease of implementation of K-means have contributed to its frequent use in applied research, sometimes without equivalent attention to its assumptions and limitations. Chen and Witten [2023] demonstrated that when groups are first created through K-means, classical post hoc comparisons such as ANOVA can produce inflated Type I error because the same data are used both to define and test the clusters. Their findings suggest that substantial between-group differences may emerge even in the absence of true population categories, precisely because clustering algorithms are designed to maximize separation in the observed data. This issue is highly relevant to psychological research, where cluster-derived groups are often subsequently compared using inferential procedures and interpreted as psychologically meaningful profiles, as seen in applied studies using clustering approaches in psychopathology and psychological assessment [Erridge et al., 2026; Orchard et al., 2024; Liu et al., 2022].

### 1.1 Objectives

The present article therefore offers a constructive methodological critique of classical and spherical K-means clustering in psychometric research. We do not argue that K-means should be avoided. On the contrary, K-means can be a valuable exploratory tool when its assumptions, limitations, and descriptive nature are made explicit. Our concern is with interpretation: the tendency to treat K-means partitions as if they were direct and even better evidence of psychological phenotypes or naturally occurring profiles. The analyses are organized around the following questions:

1. Can K-means produce interpretable clusters even when no true group structure exists?
2. Does classical K-means transform continuous psychometric variation into artificial categories?
3. How does correlation among psychological variables affect K-means solutions?
4. Does spherical K-means reveal profile-shape differences that are not captured by classical K-means?
5. How should an “optimal” value of $k$ be interpreted when silhouette values are low?

## 2 Methodology

This study was designed to examine how K-means-based clustering behaves when applied to psychological data with different underlying structures. To examine this issue, we combine controlled simulations with real psychometric data. The simulations include random data, ideal Gaussian data, multimodal Gaussian data, and correlated Gaussian data designed to approximate common structures in psychological measurement. We then evaluate whether similar partitioning patterns emerge in the SMARVUS dataset, using its large, multilingual, cross-cultural structure as an empirical test case.

Finally, to contrast these psychometric scenarios with a setting in which clustering is more naturally expected, we also simulated a flow-cytometry-like dataset, approximating the everyday analytical context of biomedical laboratory work, where distinct cell populations are often visually and statistically identifiable.

## 2.1 Data Analysis

All analyses were conducted in Python using the *scikit-learn* library [Pedregosa et al., 2011]. Classical K-means was used to assess partitions based on Euclidean proximity and overall score magnitude, whereas spherical K-means was used as a contrastive approach emphasizing angular similarity and relative response profiles [Hornik et al., 2012; Dhillon and Modha, 2001]. For both methods, the selected features were standardized using *StandardScaler*, and no missing values were present.

Candidate solutions were explored across a range of $k$ values, typically from 2 to 10. For each value of $k$, the algorithm was initialized multiple times using K-means++, and the best solution was retained according to the optimization criterion. This procedure was adopted because K-means is sensitive to initial starting points and may converge to local rather than global optima, a problem highlighted by Steinley [2006].

Three quantitative criteria were used to evaluate the clustering results: the silhouette coefficient, inertia, and the adjusted Rand index (ARI). The silhouette coefficient was used to guide the choice of $k$ by estimating how well each observation fit within its assigned cluster relative to the nearest alternative solution. Higher silhouette values indicate more compact and better-separated clusters, whereas low values suggest weak or ambiguous structure, even when a mathematical partition can still be produced [Gao et al., 2023; Steinley, 2006].

Inertia, or within-cluster sum of squares (SSE), was also examined through the elbow method to evaluate whether increasing $k$ produced a meaningful reduction in within-cluster dispersion or merely incremental gains from subdividing the data. Finally, the ARI was used to assess reproducibility across repeated runs by quantifying the agreement between clustering partitions obtained from different random initializations of the same data and the same value of $k$. ARI values close to 1 indicate highly consistent partitions, whereas lower values indicate that the solution depends more strongly on initialization and may therefore be less stable [Gao et al., 2023; Steinley, 2006].

In addition to the quantitative criteria, Principal Component Analysis (PCA) projections and cluster plots were used as visual diagnostic tools throughout the project. PCA was applied to reduce the original multidimensional feature space to two or three dimensions, allowing the clustering solutions to be inspected graphically. PCA was used only for visualization and not as a feature-extraction step for clustering. This choice was made because PCA emphasizes variance rather than cluster structure, and recent evidence shows that clustering on principal components can be problematic, particularly when subgroup structure is nonlinear or not linearly separable [Lötsch & Ultsch, 2024].

## 2.2 Datasets

Five synthetic datasets and one empirical dataset were analyzed. Each simulated dataset contained 2,000 observations and 6 continuous variables. The same dimensionality was used across simulations to ensure that differences in clustering behavior reflected the structure of the data rather than sample size or number of features in the model space.

### 2.2.1 Dataset 1: Random Data

The first dataset consisted of fully random data generated without any latent group structure (Figure 1). The purpose of this condition was to demonstrate that clustering algorithms, especially K-means, can always return a partition even when no meaningful structure exists in the data. If K-means produces apparently interpretable centroids in random data, this illustrates that the existence of a clustering solution is not, by itself, evidence of psychological subgroups.

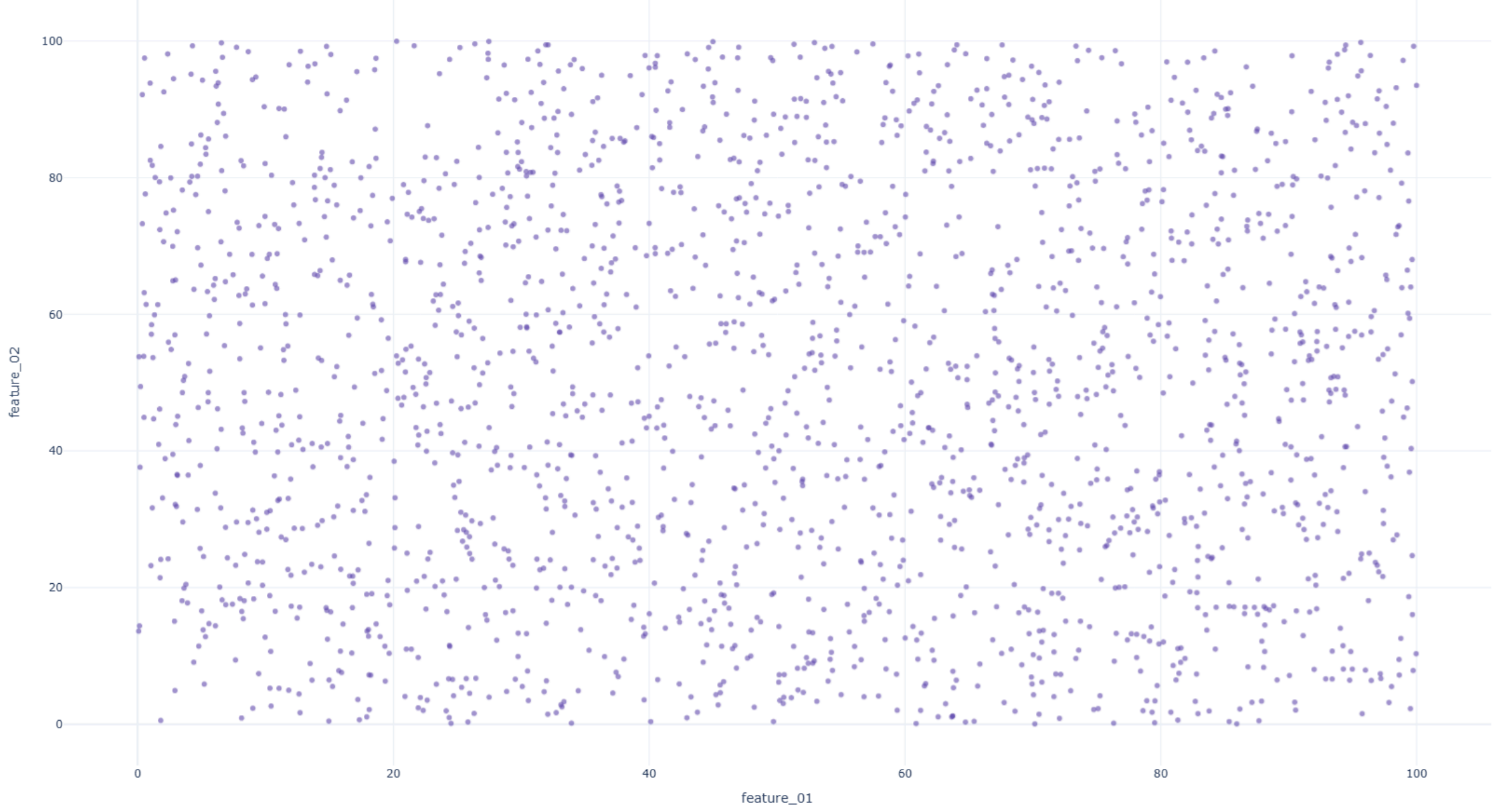

**Figure 1.** Pairwise comparison of features 1 and 2 in dataset 1.

### 2.2.2 Dataset 2: Unimodal Gaussian Data

The second dataset consisted of variables generated from an ideal Gaussian distribution, but with no discrete latent classes (Figure 2). This condition was designed to mimic a common psychometric situation: individuals vary continuously along psychological dimensions, but there are no psychological associations between the constructs measured.

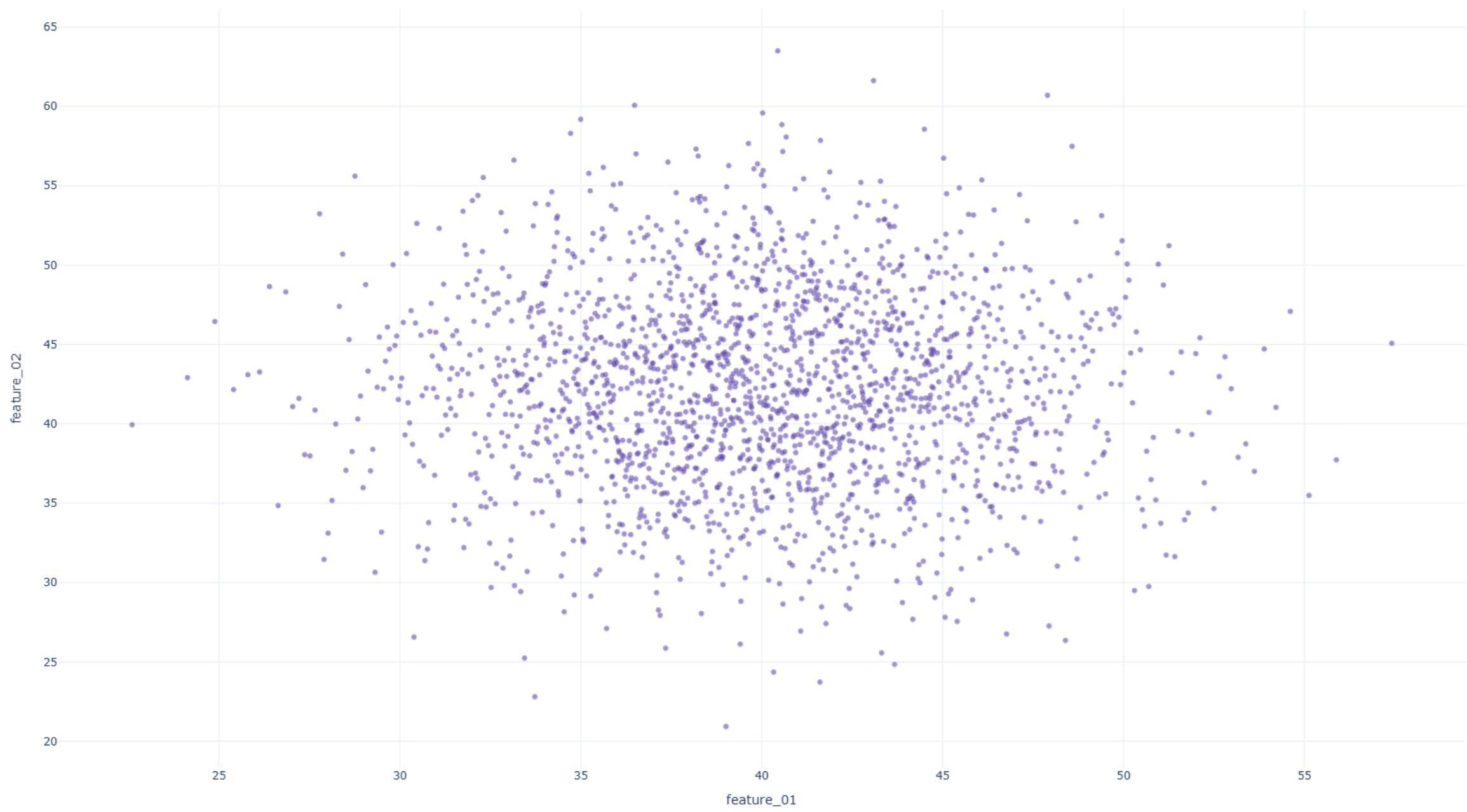

**Figure 2.** Pairwise comparison of features 1 and 2 in dataset 2.

### 2.2.3 Dataset 3: Unimodal Correlated Gaussian Data

The third dataset consisted of correlated Gaussian variables. This condition was designed to approximate a common structure in psychological measurement, where features are often intercorrelated (Figure 3).

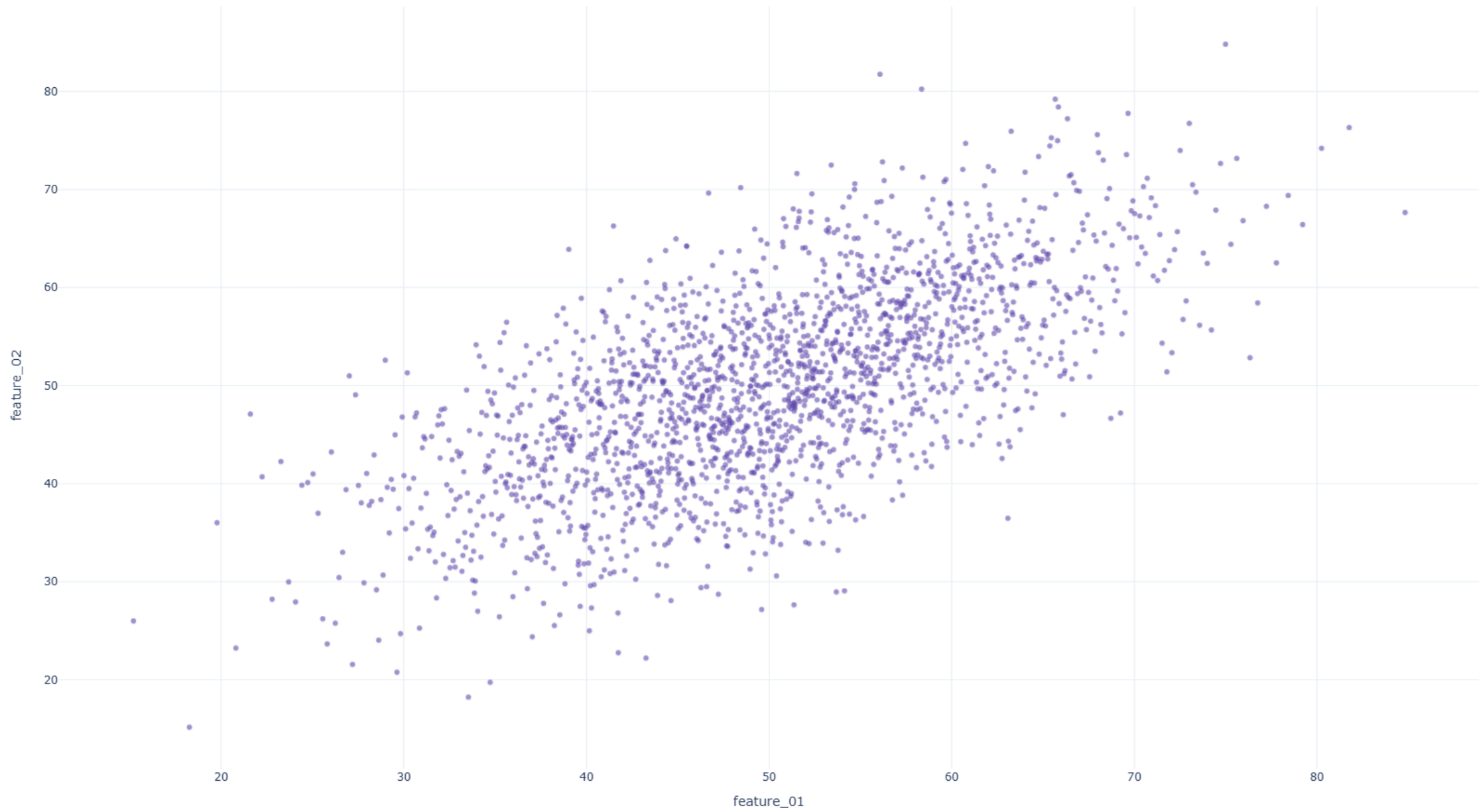

**Figure 3.** Pairwise comparison of features 1 and 2 in dataset 3.

### 2.2.4 Dataset 4: Multimodal Gaussian Data

The fourth dataset consisted of data generated from multiple Gaussian distributions. This condition served as a positive control ($k$ = 5) because it represents an idealized case in which true group structure is present and approximately compatible with the assumptions of classical K-means (Figure 4).

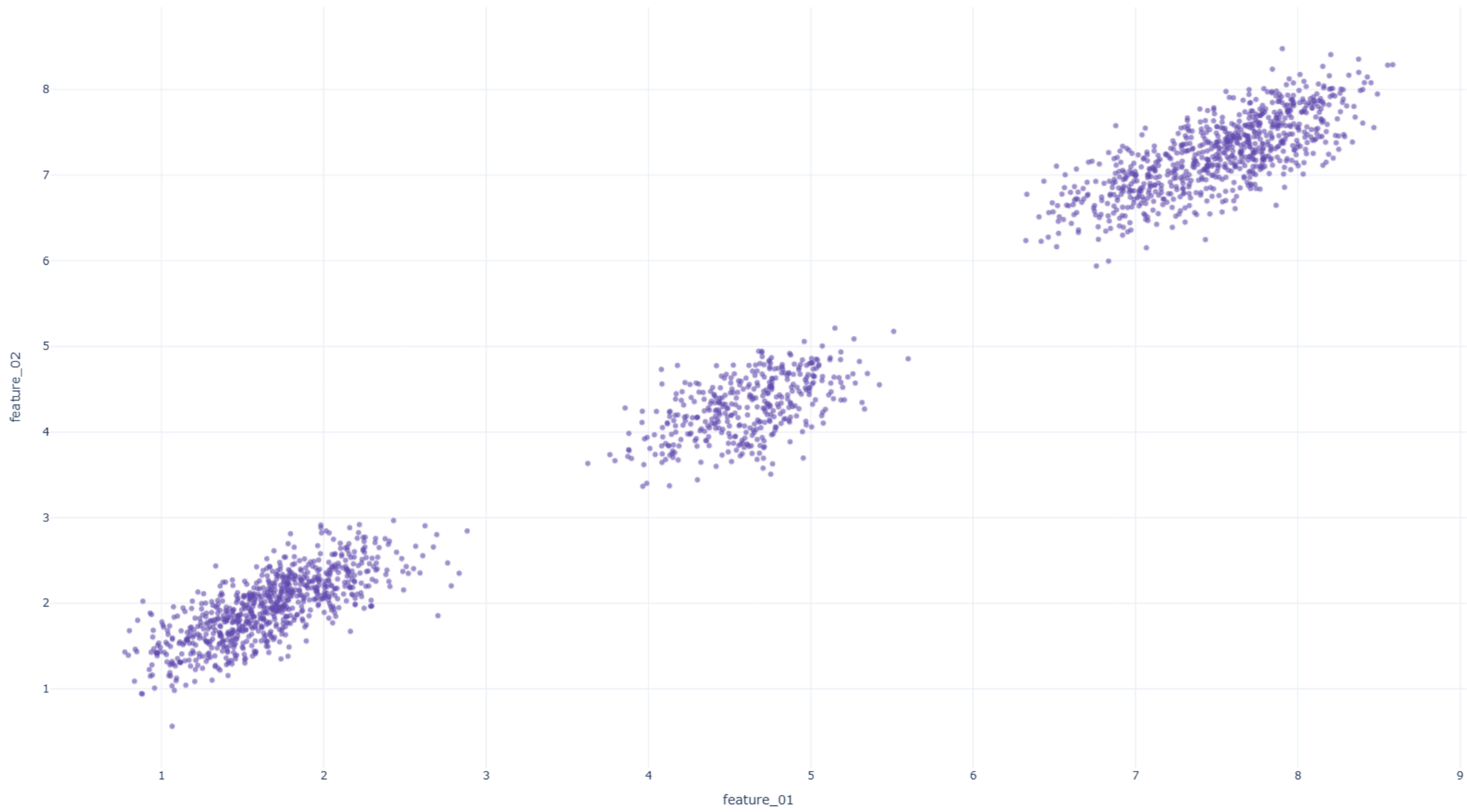

**Figure 4.** Pairwise comparison of features 1 and 2 in dataset 4.

### 2.2.5 Dataset 5: Cytometer-like Data

For the cytometer-like dataset, we also simulated a multivariate structure intended to represent a condition in which clustering is substantively expected, resembling the type of population structure commonly observed in flow-cytometry data (Figure 5).

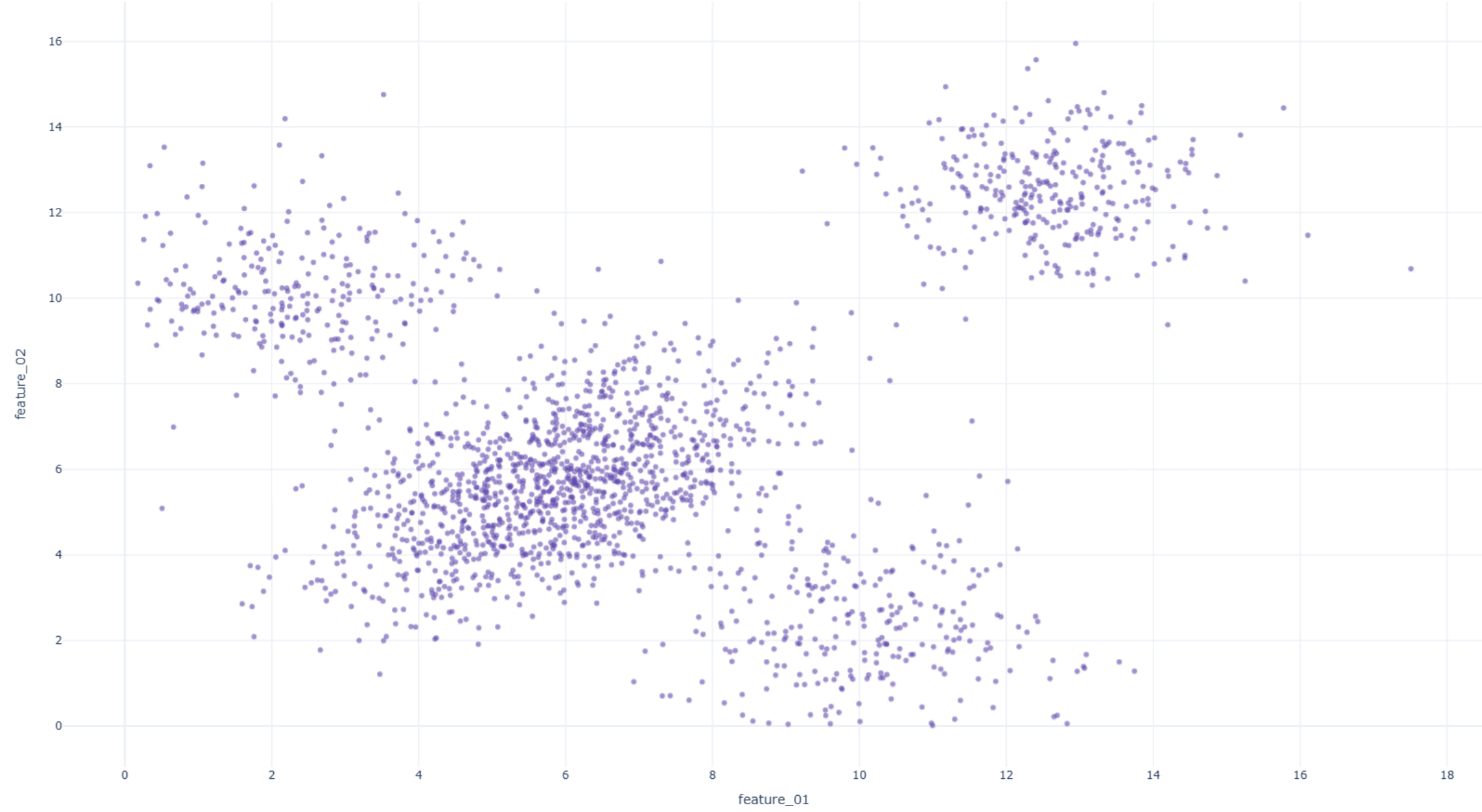

**Figure 5.** Pairwise comparison of features 1 and 2 in dataset 5.

Small amounts of noise and partial overlap were intentionally introduced to avoid an artificially perfect separation and to approximate the visual appearance of empirical cytometry-like measurements, in which cell populations may be distinguishable but not completely isolated. There are five types of cells present in the dataset.

#### 2.2.6 Dataset 6: Empirical Data

The empirical component of the study uses the SMARVUS dataset, an international multicenter psychometric dataset containing survey responses from 12,570 university students from 100 universities, 35 countries, and 21 languages. Data cleaning was performed sequentially. Participants were excluded if they failed any embedded attention-check. Additional exclusions were applied to participants who failed the attention-amnesty check. This procedure reduced the dataset from 12,570 to 8,360 participants (Figure 6).

The final analytical sample was predominantly composed of women. Among participants with available gender information, 6,727 identified as female/woman, 1509 as male/man, and 81 as another gender. The sample was also concentrated in younger age ranges: 6,166 participants were aged 18–21 years, 1,450 were aged 22–25 years, and 494 were aged 26 years or older.

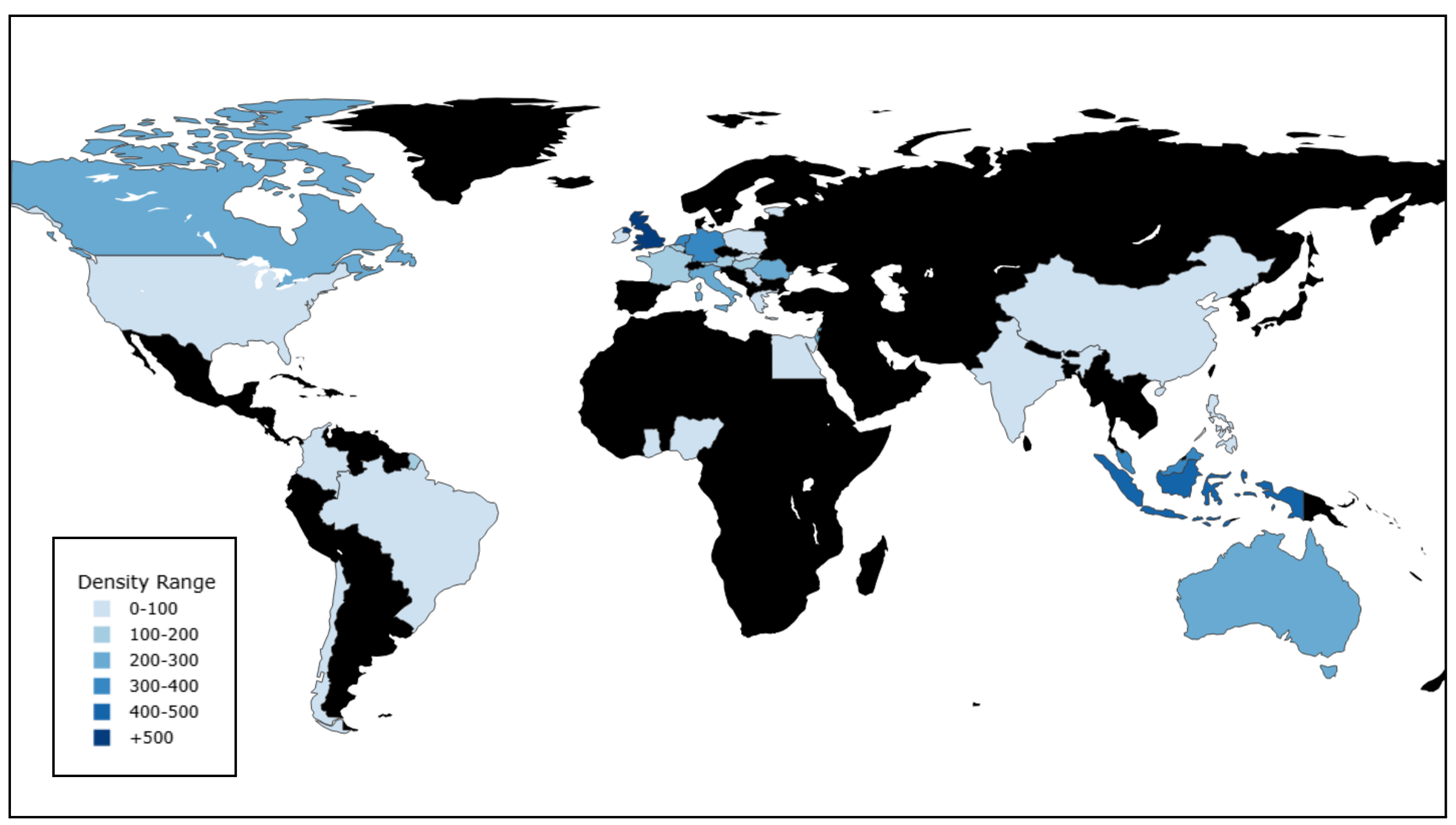

**Figure 6.** Participants density around the world.

Following the same feature-dimensionality used in the other controlled examples, six psychometric features were retained in the empirical model space: STARS, R-TAS, BFNE-S, LSAS-SR, CAS, and IUS-SF (Figure 7). These features were selected to provide a compact and interpretable psychological space for comparing whether K-means partitions reflected meaningful profiles or broader dimensional gradients:

1. **The Statistics Anxiety Rating Scale (STARS)**, developed by Cruise et al. [1985], is a self-report psychometric instrument designed to assess individuals' anxiety related to statistics, particularly in academic contexts. It comprises multiple dimensions capturing both emotional reactions to statistical tasks and attitudes toward statistics, although many studies distinguish between its anxiety-specific and attitudinal components [Hanna et al., 2008; Papousek et al., 2012].

2. **The Revised Test Anxiety Scale (R-TAS)**, developed by Benson and El-Zahhar [1994], is a self-report measure designed to assess test anxiety as a multidimensional construct, comprising four correlated dimensions: worry, tension, bodily symptoms, and test-irrelevant thinking.
3. **The Brief Fear of Negative Evaluation Scale – Straightforward Items (BFNE-S)**, derived from the original BFNE [Leary, 1983], is a self-report measure designed to assess fear of negative evaluation, a core component of social anxiety [Rodebaugh et al., 2004].
4. **The Liebowitz Social Anxiety Scale – Self-Report (LSAS-SR)** is a self-report instrument evaluated by Baker et al. [2002], demonstrating good reliability and validity for assessing social anxiety, although its factor structure remains debated. It is based on the original clinician-administered LSAS developed by Michael R. Liebowitz [1987], which served as the foundation for its structure and content.
5. **The Creativity Anxiety Scale (CAS)**, developed by Daker et al. [2019], is a self-report instrument designed to assess anxiety specifically associated with engaging in creative thinking across domains.
6. **The Intolerance of Uncertainty Scale – Short Version (IUS-SF)**, developed by Carleton et al. [2007] as a reduced form of the original scale by Freeston et al. [1994], is a self-report psychometric instrument designed to assess individuals' tendency to perceive uncertain situations as distressing and unacceptable.

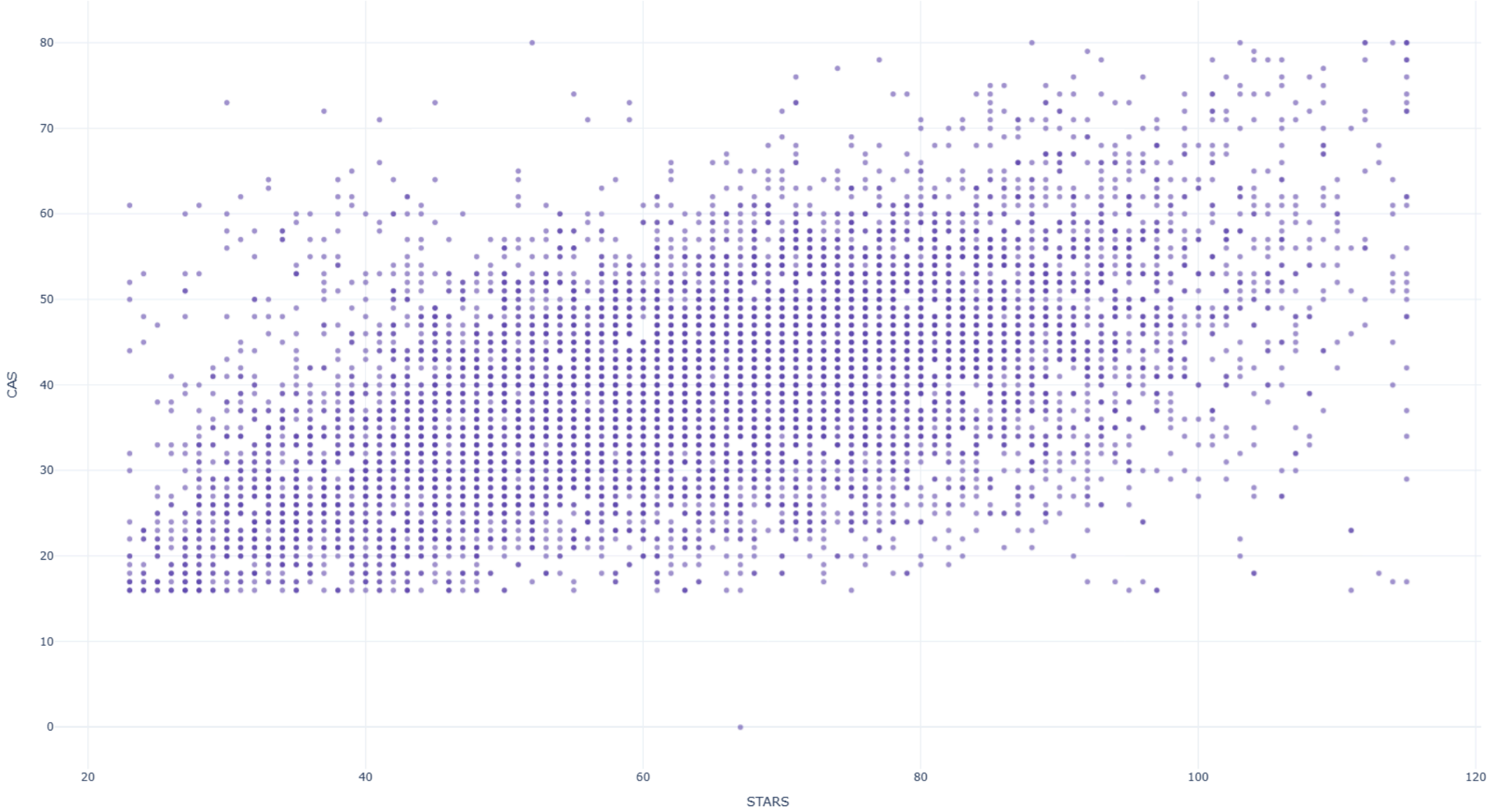

**Figure 7.** Pairwise comparison of STARS and CAS in SMARVUS dataset.

## 3 Results

Datasets 1, 2, 3, 4, and the cytometer-like dataset generated and used in the simulations are publicly available in the project repository, as are details of the analyses performed:

https://osf.io/dtbxw/overview?view_only=b040b4a9458b43439cdc189a5745d3f9

### 3.1 Clustering Random Data

For the random-data condition, the choice of $k$ illustrates the ambiguity of cluster selection when no true underlying structure is present (Figure 8). The silhouette curve shows uniformly low values across all tested solutions, ranging only around 0.12–0.135. Although the highest silhouette value occurs at $k$ = 10, this should not be interpreted as evidence for ten meaningful clusters. The increase is very small and likely reflects the tendency of K-means to impose increasingly fine geometric partitions on noise.

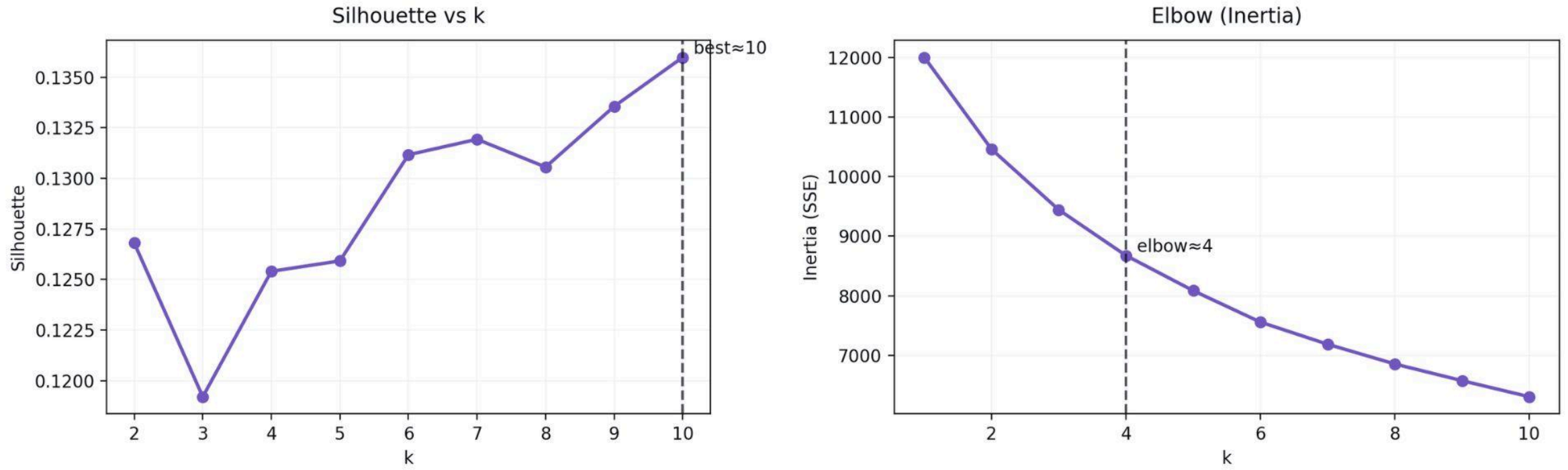

**Figure 8.** Silhouette and SSE vs $k$ for random data.

The elbow plot suggests a possible bend around $k = 4$, where the reduction in inertia begins to slow. However, this elbow is also weak and gradual rather than sharply defined. This indicates that additional clusters continue to reduce within-cluster variance simply because K-means is subdividing the feature space, not because the data contain well-separated groups. Thus, in random data, neither criterion provides strong evidence for a substantively meaningful cluster solution.

However, the profile plot itself does not provide clear evidence that the clustering solution is invalid (Figure 9). Visually, the algorithm still produces differentiated average profiles across clusters, which can easily create the impression of meaningful subgroup structure. The main indication that the solution may not reflect stable underlying groups emerges only from the reproducibility analysis. Specifically, the ARI across different K-means initializations was moderate (ARI = 0.330, SD = 0.061), indicating that different random starts frequently produced different partitions for the same dataset.

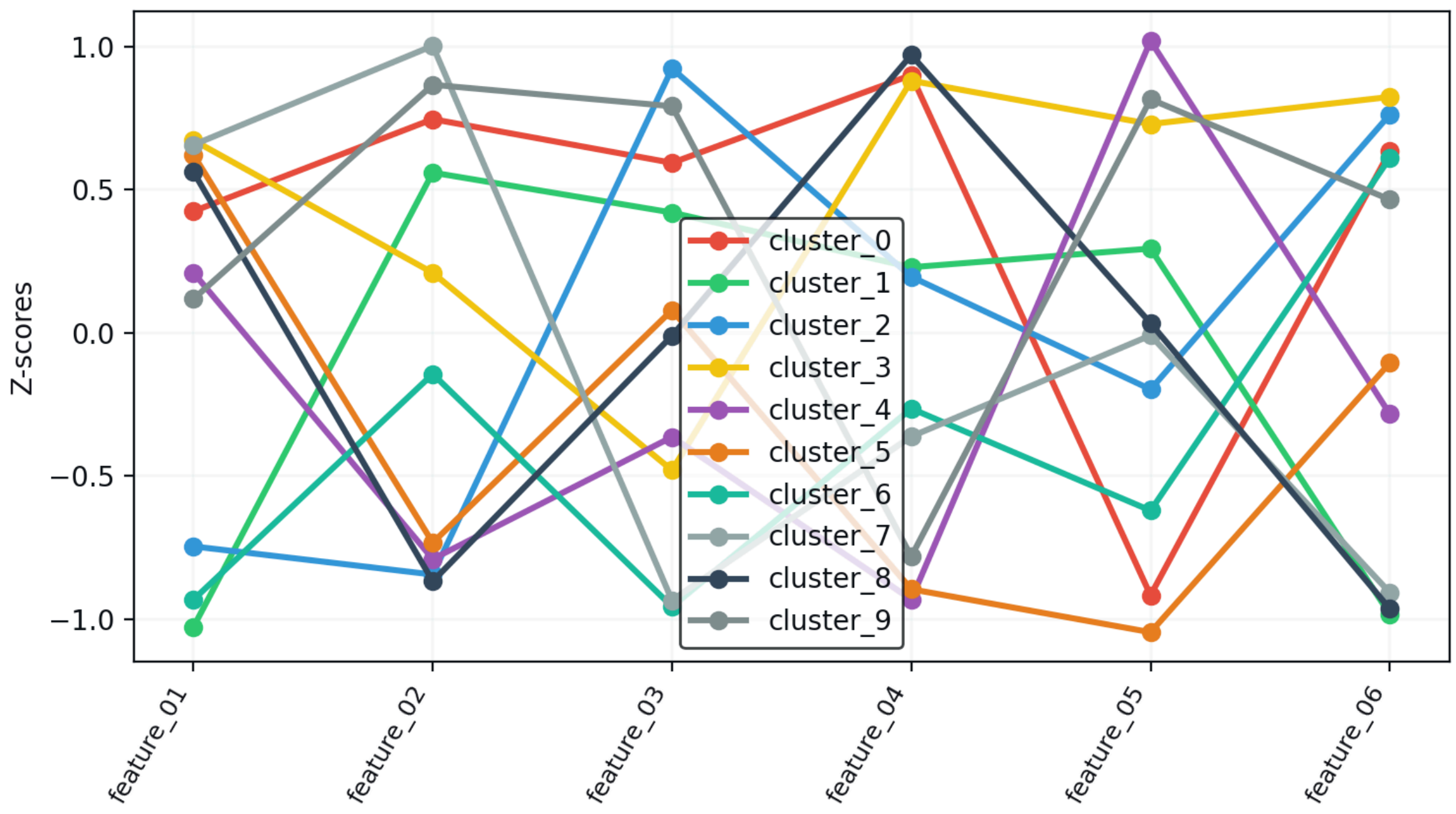

**Figure 9.** Cluster profile plot for random data ($k$ = 10).

In addition to classical K-means, we applied spherical K-means to the random-data condition. Even in random data, spherical K-means produced a visually structured multi-cluster solution (Figure 10).

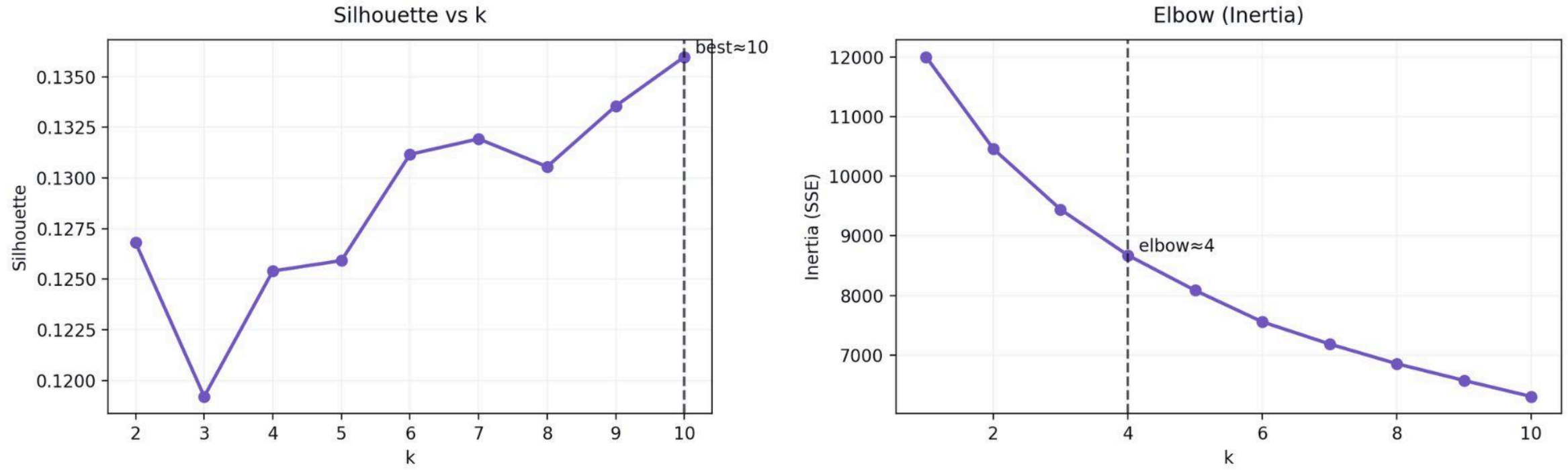

**Figure 10.** Silhouette (cosine) and SSE (L2) vs $k$ for random data.

In the 2D PCA projection, the clusters appear distributed across the projected space, but with substantial overlap and no clearly isolated regions, suggesting that the algorithm imposed angular partitions on a continuous cloud (Figure 11). This interpretation is supported by the stability analysis, indicating that different random starts produced partially inconsistent partitions for the same data (ARI = 0.342, SD = 0.075).

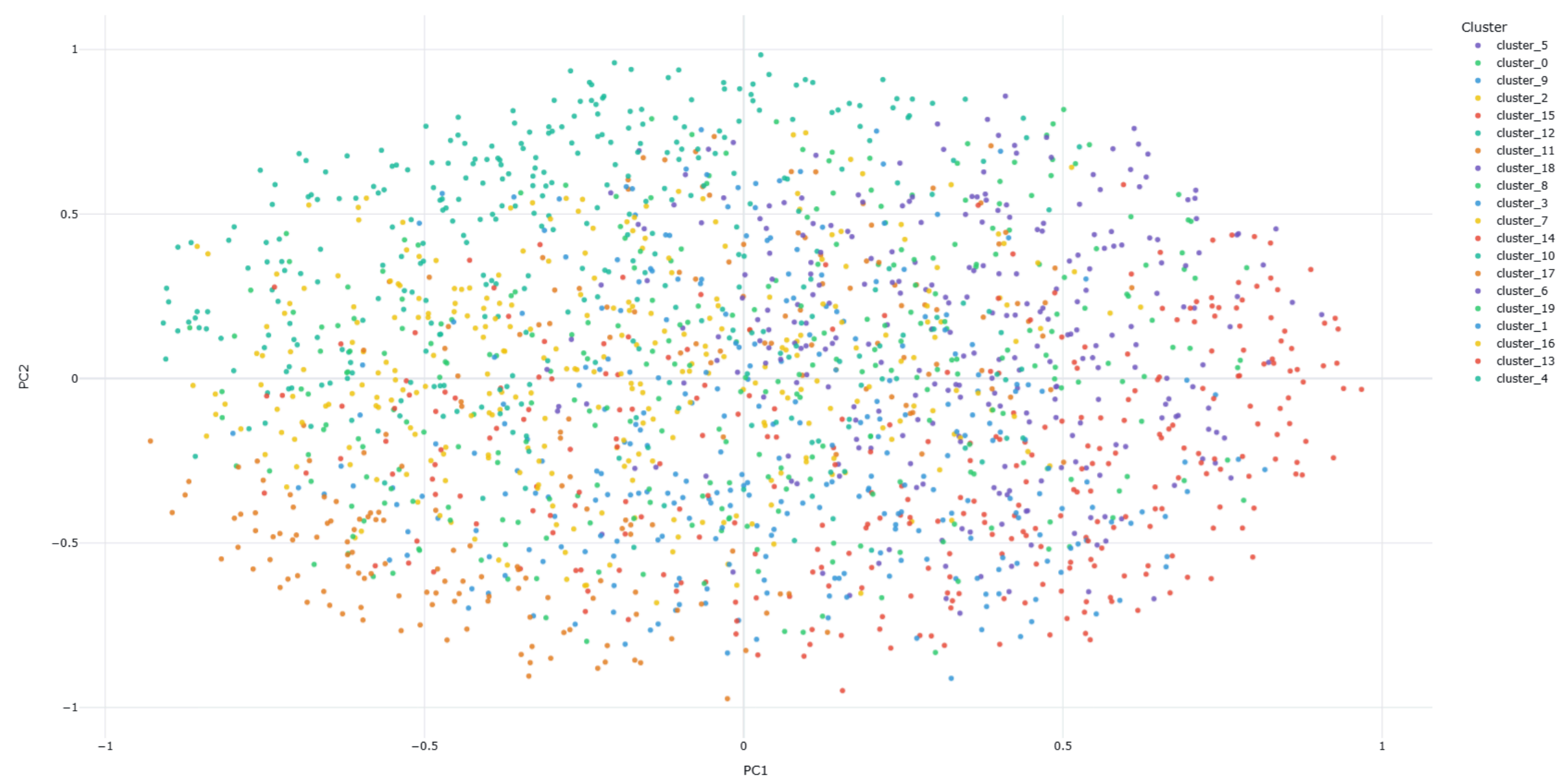

**Figure 11.** 2D PCA projection for random data ($k = 10$).

### 3.2 Clustering Unimodal Gaussian Data

The second simulation used unimodal Gaussian data. As in the random-data condition, K-means still returned a cluster solution (Figure 12). Classical K-means selected $k = 7$ according to the silhouette criterion. However, the absolute silhouette value was extremely low, reaching only approximately 0.115 with no plateau, which indicates very weak cluster separation. The inertia curve was similarly inconclusive.

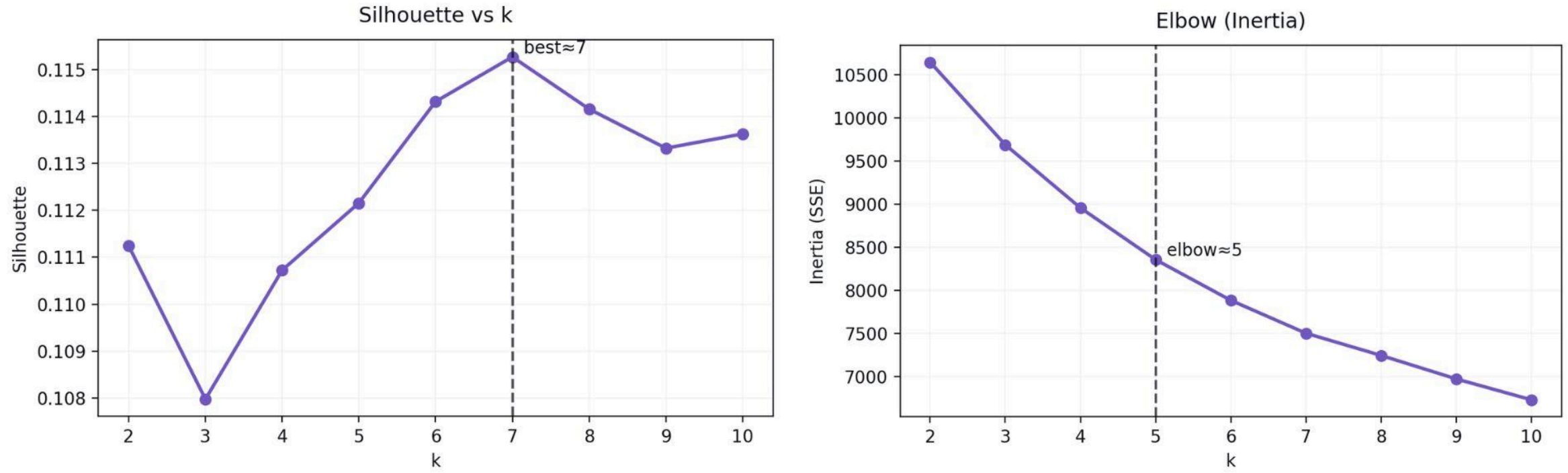

**Figure 12.** Silhouette and SSE vs $k$ for unimodal Gaussian data.

As in the random-data condition, the cluster-profile plot alone does not provide a clear indication that the clustering solution is unstable (Figure 13). Although the profiles appear differentiated across features, such visual separation can still create the impression of meaningful subgroup structure. The instability of the solution only becomes evident when examining the reproducibility analysis (ARI = 0.343, SD = 0.133), indicating that different random starts frequently generated different partitions for the same dataset.

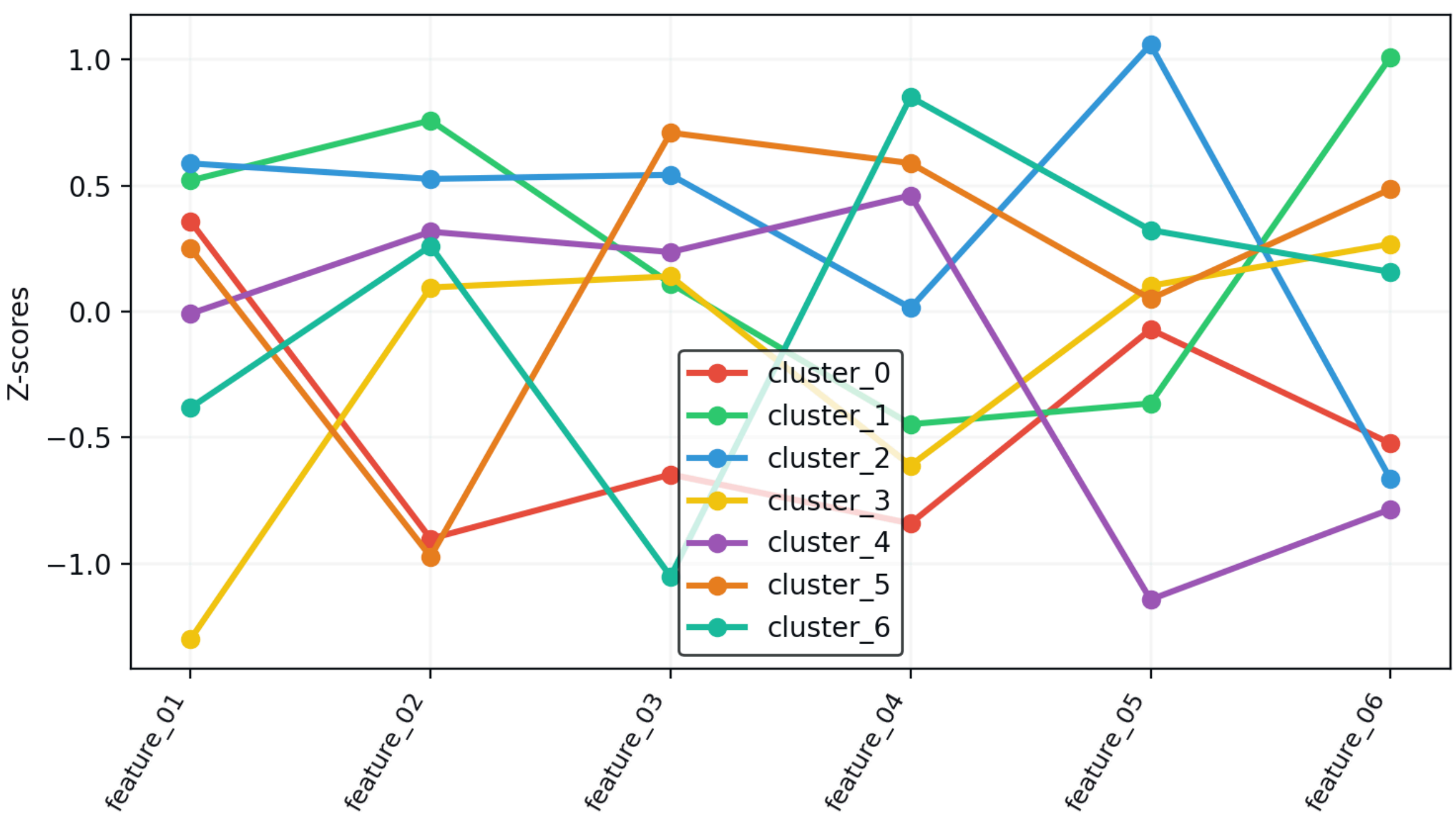

**Figure 13.** Cluster-profile plot for unimodal Gaussian data ($k = 7$).

Spherical K-means was also applied to the same Gaussian dataset. In this case, the cosine silhouette curve continued to increase across much of the tested range and selected approximately $k = 7$ as the best solution, with a maximum cosine silhouette around 0.23 (Figure 14).

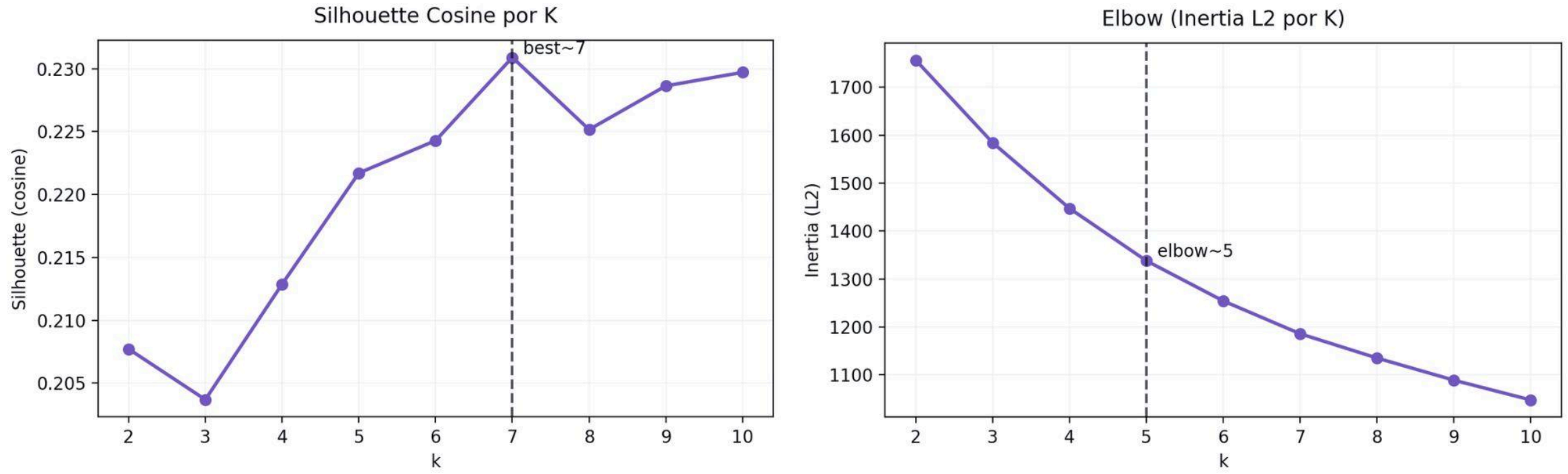

**Figure 14.** Silhouette (cosine) and SSE (L2) vs $k$ for unimodal Gaussian data.

Although this value is higher than the classical Euclidean silhouette, the PCA projection still shows extensive overlap among clusters, with colored points dispersed throughout the same continuous cloud rather than forming clearly isolated regions (Figure 15). Thus, the spherical solution appears to capture angular subdivisions of the Gaussian distribution rather than meaningful cluster structure. The stability analysis again supports this interpretation (ARI = 0.401, SD = 0.101).

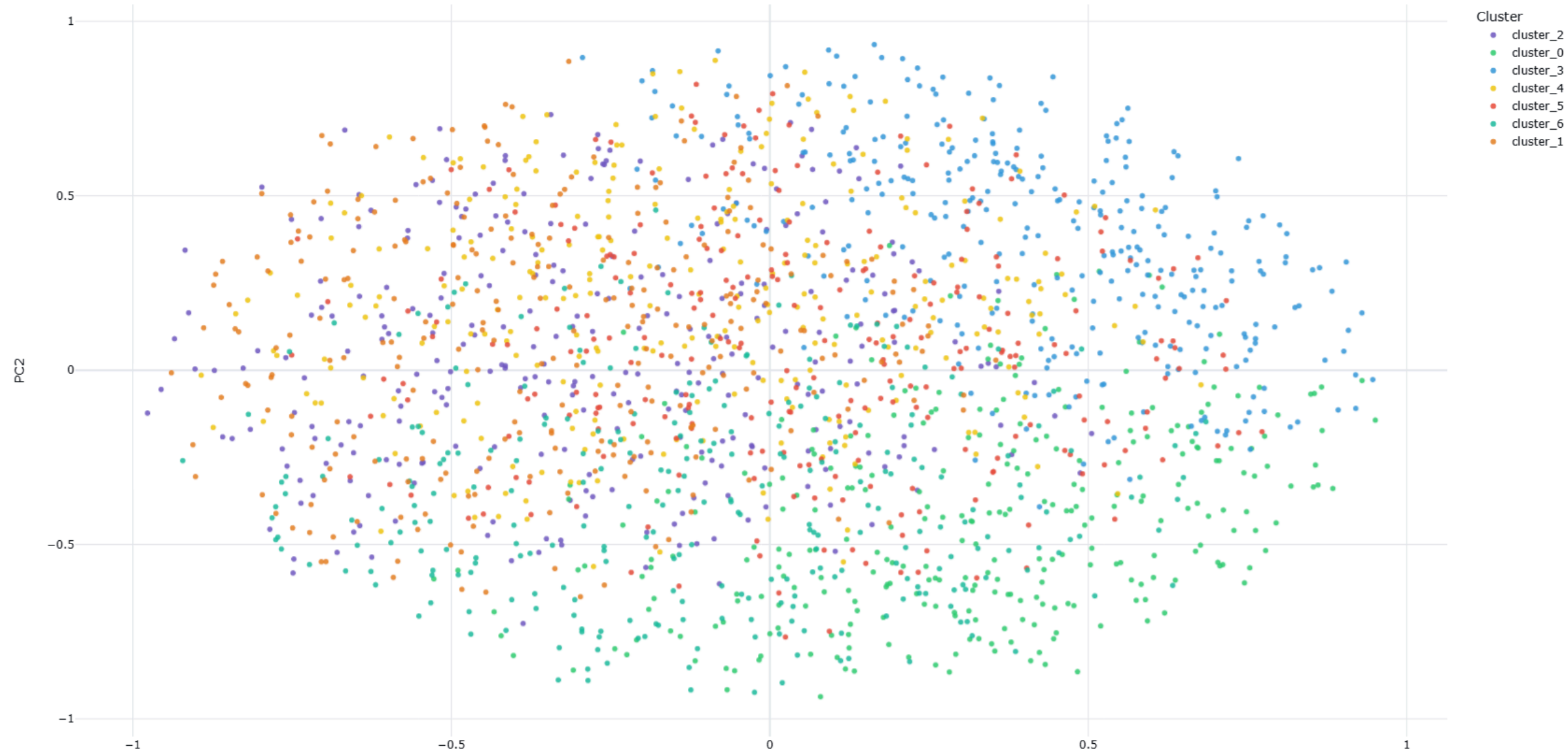

**Figure 15.** 2D PCA projection for unimodal Gaussian data ($k = 7$).

Importantly, this value was comparable to the cosine silhouette observed in the random-data condition, indicating that the apparent improvement in visual separation does not necessarily reflect a stronger or more meaningful latent structure. Rather, spherical K-means appears to impose a directional partition on the Gaussian space, producing a visually coherent division even when the underlying data-generating process remains continuous and unimodal.

### 3.3 Clustering Unimodal Correlated Gaussian Data

For the unimodal correlated Gaussian dataset, the application of K-means clustering with $k = 2$ (best) resulted in a highly stable partitioning of the data (Figure 16). The Adjusted Rand Index (ARI) reached 1.00 with a standard deviation of 0.00 across runs, indicating complete consistency in the clustering solution regardless of initialization. The average silhouette value was slightly above 0.26, suggesting only modest separation between clusters despite the stability of results.

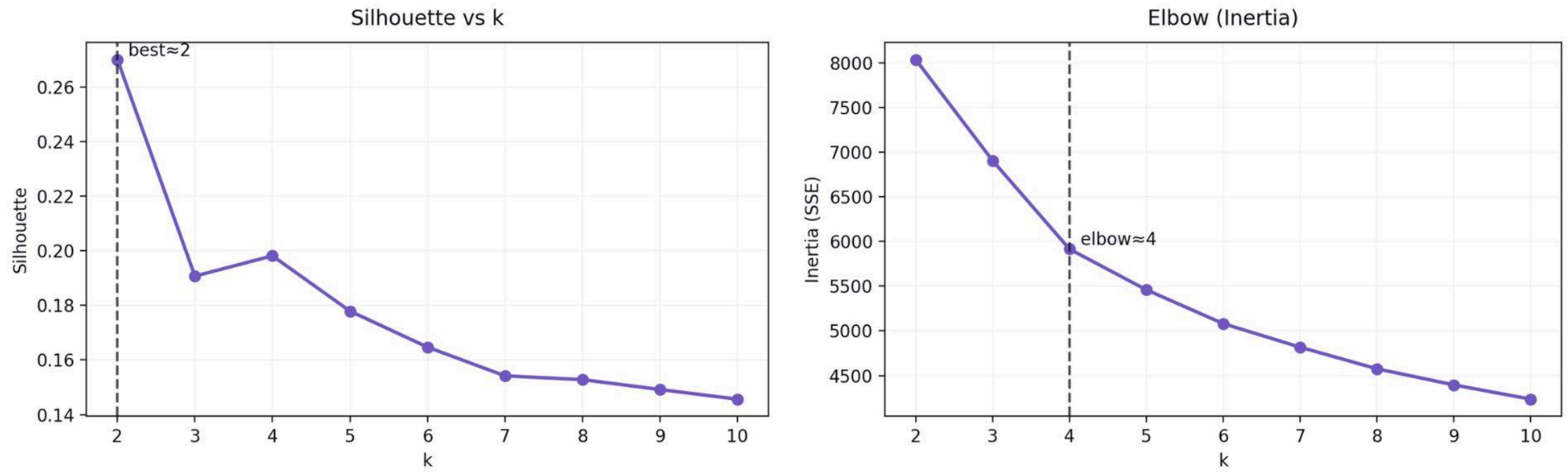

**Figure 16.** Silhouette and SSE vs k for unimodal correlated Gaussian data.

Given that the data were generated from unimodal Gaussian distribution with correlated features, the algorithm effectively split an elongated, continuous structure into two elongated halves (Figure 17). Thus, the identified clusters should be interpreted not as meaningful subpopulations, but as an artificial partition of a continuous, anisotropic distribution.

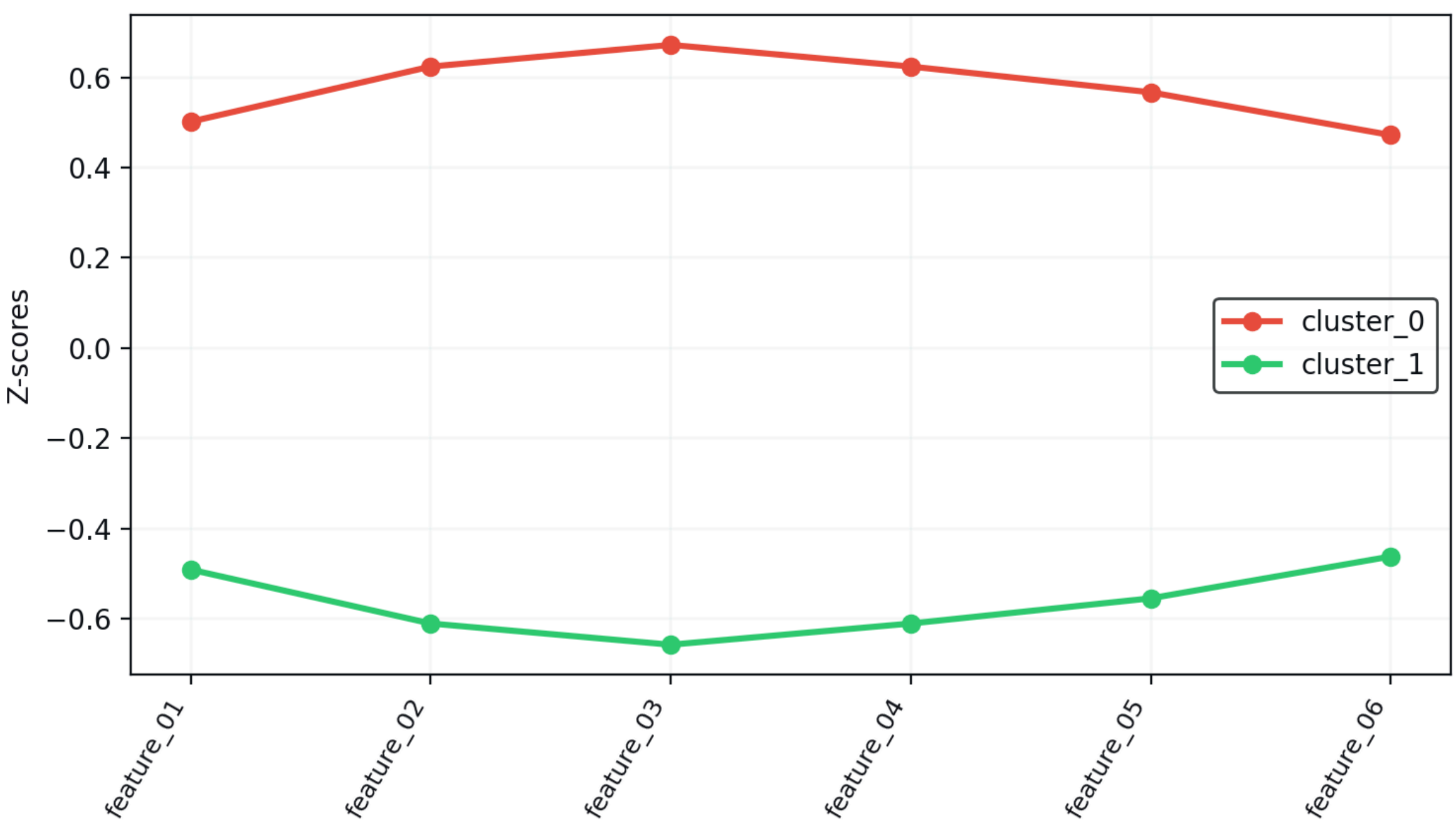

**Figure 17.** Profile-plot for unimodal correlated Gaussian data ($k = 2$).

The application of spherical K-means clustering with $k = 2$ also produced a highly stable partitioning of the data (Figure 18). The Adjusted Rand Index (ARI) reached 0.999, with a standard deviation of 0.001 across runs, indicating that the solution was almost perfectly reproducible despite random initialization. The cosine silhouette analysis also selected $k = 2$ as the best solution, with a silhouette value around 0.46, which is higher than the value obtained with classical K-means.

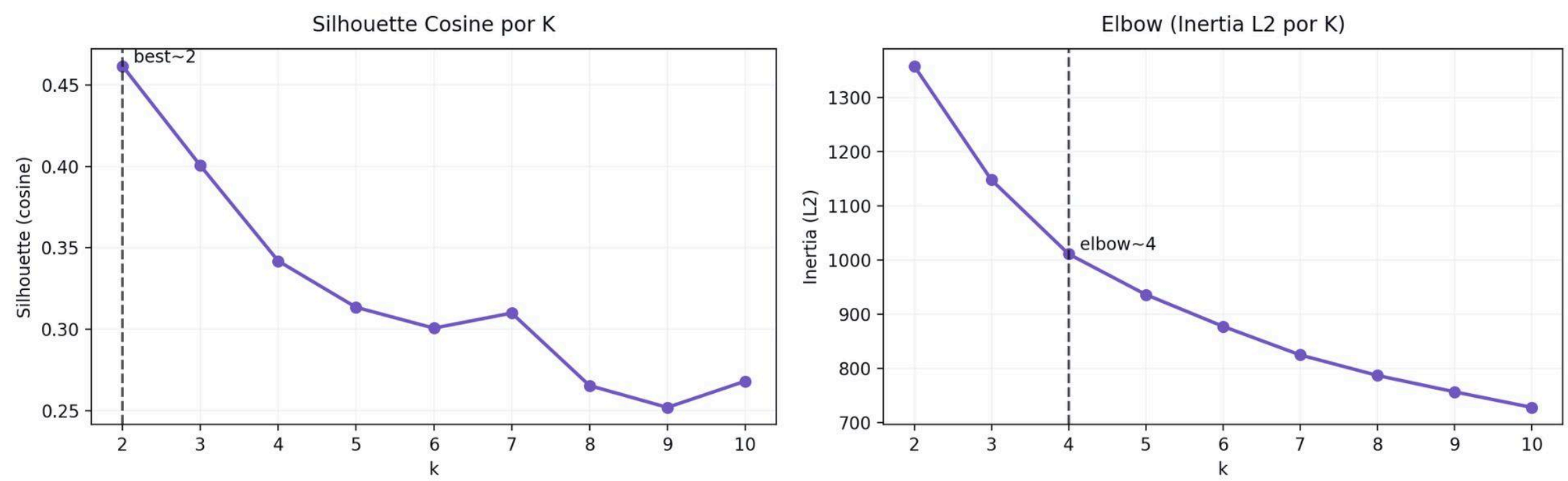

**Figure 18.** Silhouette (cosine) and SSE (L2) vs k for unimodal correlated Gaussian data.

However, this result should not be interpreted as evidence for two genuine psychological phenotypes. In the PCA projection, this produced a clear left-versus-right division of the same continuous elliptical structure, almost as if the Gaussian cloud had been cut into two hemispheres (Figure 19). Therefore, although the solution was highly stable and showed a better cosine silhouette than classical K-means, it still reflects an artificial geometric partition imposed by the algorithm rather than the discovery of meaningful subpopulations.

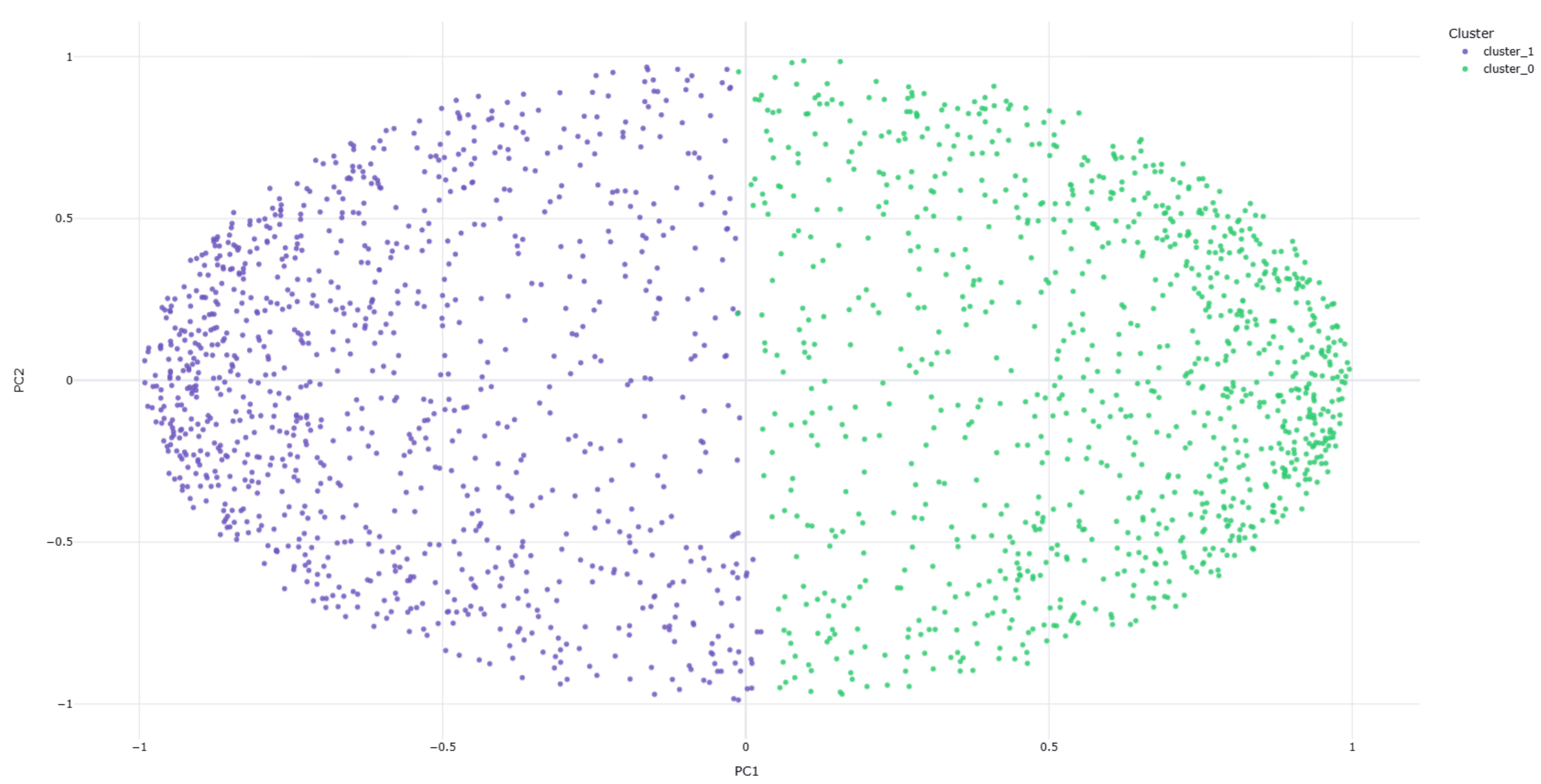

**Figure 19.** 2D PCA projection for unimodal correlated Gaussian data ($k = 2$).

### 3.4 Clustering Multimodal Gaussian Data

For the multimodal Gaussian dataset, both criteria supported $k = 5$ (Figure 20). The silhouette reached its highest value around 0.82, indicating compact and well-separated clusters. The elbow plot showed a sharp reduction in inertia up to $k = 5$, followed by a clear plateau, suggesting that additional clusters produced little improvement. Therefore, unlike the correlated Gaussian-data condition, this dataset showed a clear clustering structure, with five distinguishable groups providing the most defensible K-means solution.

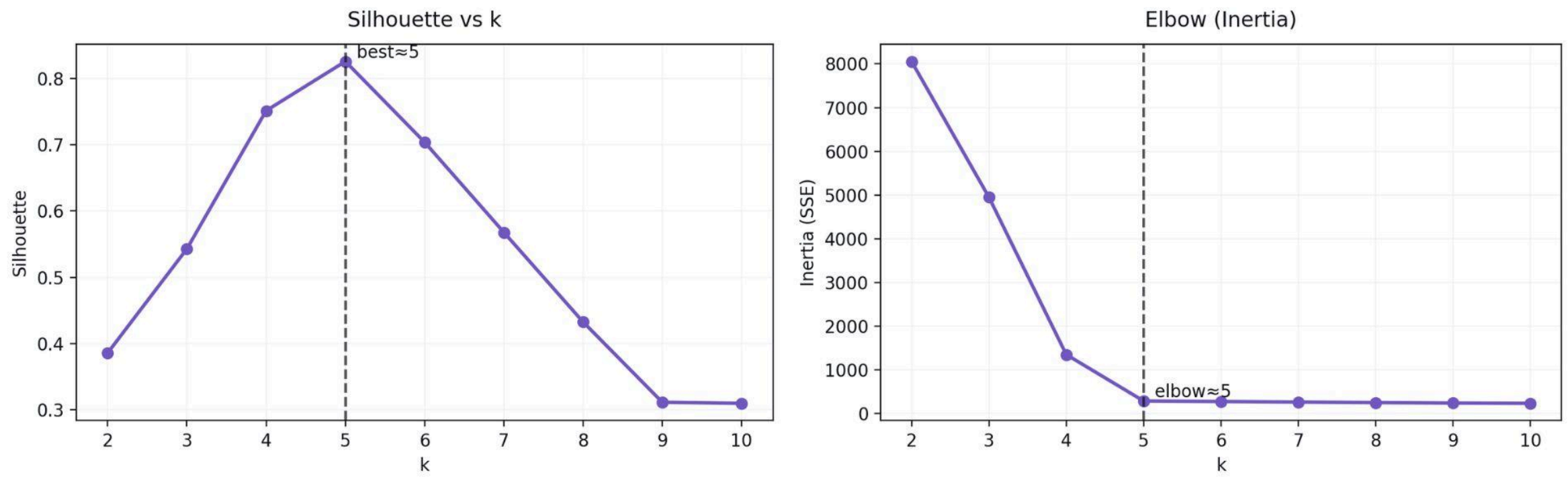

**Figure 20.** Silhouette and SSE vs k for multimodal Gaussian data.

Nevertheless, the cluster-profile plot for the multimodal Gaussian dataset is visually comparable to several other clustering solutions presented throughout the study (Figure 21).

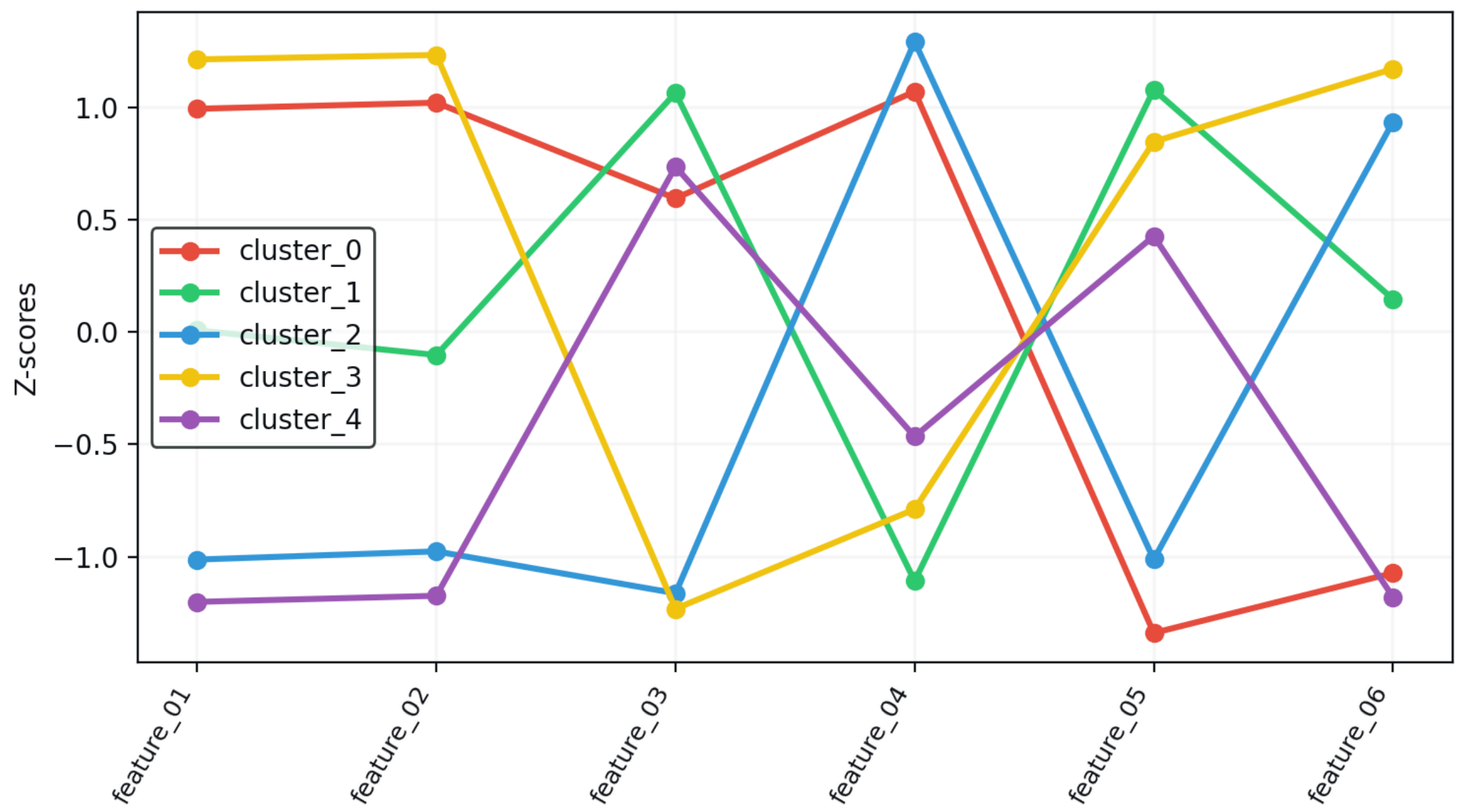

**Figure 21.** Profile-plot for multimodal Gaussian data ($k = 5$).

Spherical K-means produced the same optimal solution (Figure 22). Therefore, both classical and spherical K-means converged on the same five-cluster structure, reinforcing that this simulated dataset contained a clear multivariate clustering pattern.

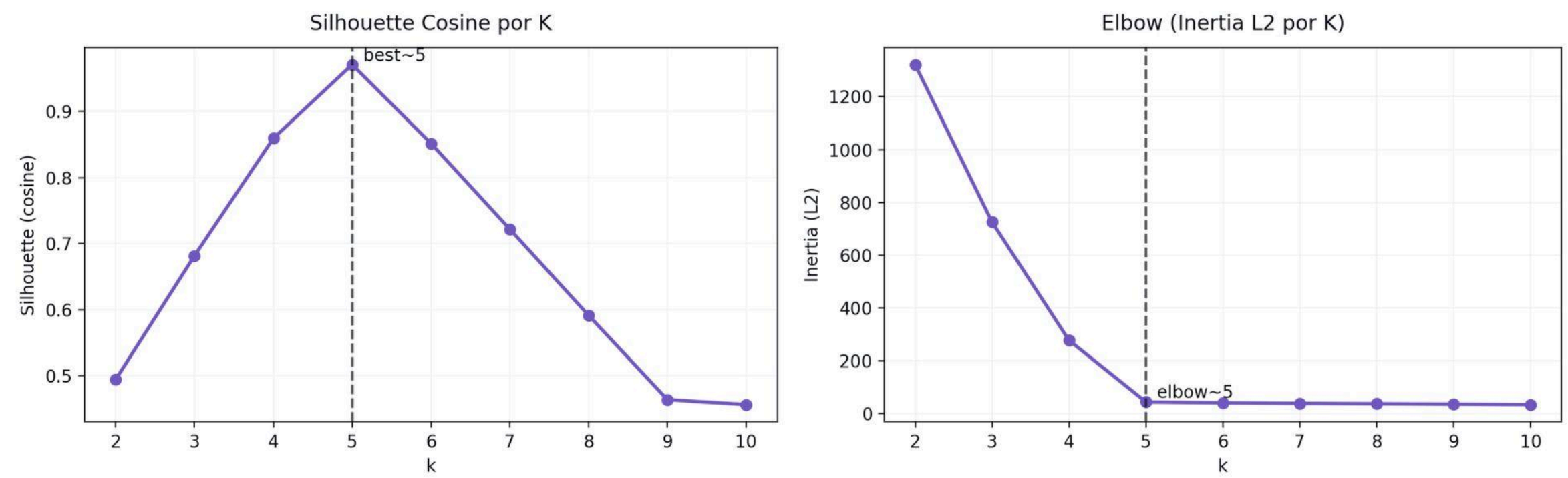

**Figure 22.** Silhouette (cosine) and SSE vs $k$ for multimodal Gaussian data.

The 3D PCA projection further reveals that this solution did not simply draw boundaries across continuous space (Figure 23). Instead, spherical K-means subdivided the data into multiple angular regions, effectively splitting the multimodal structure into smaller directional subclusters.

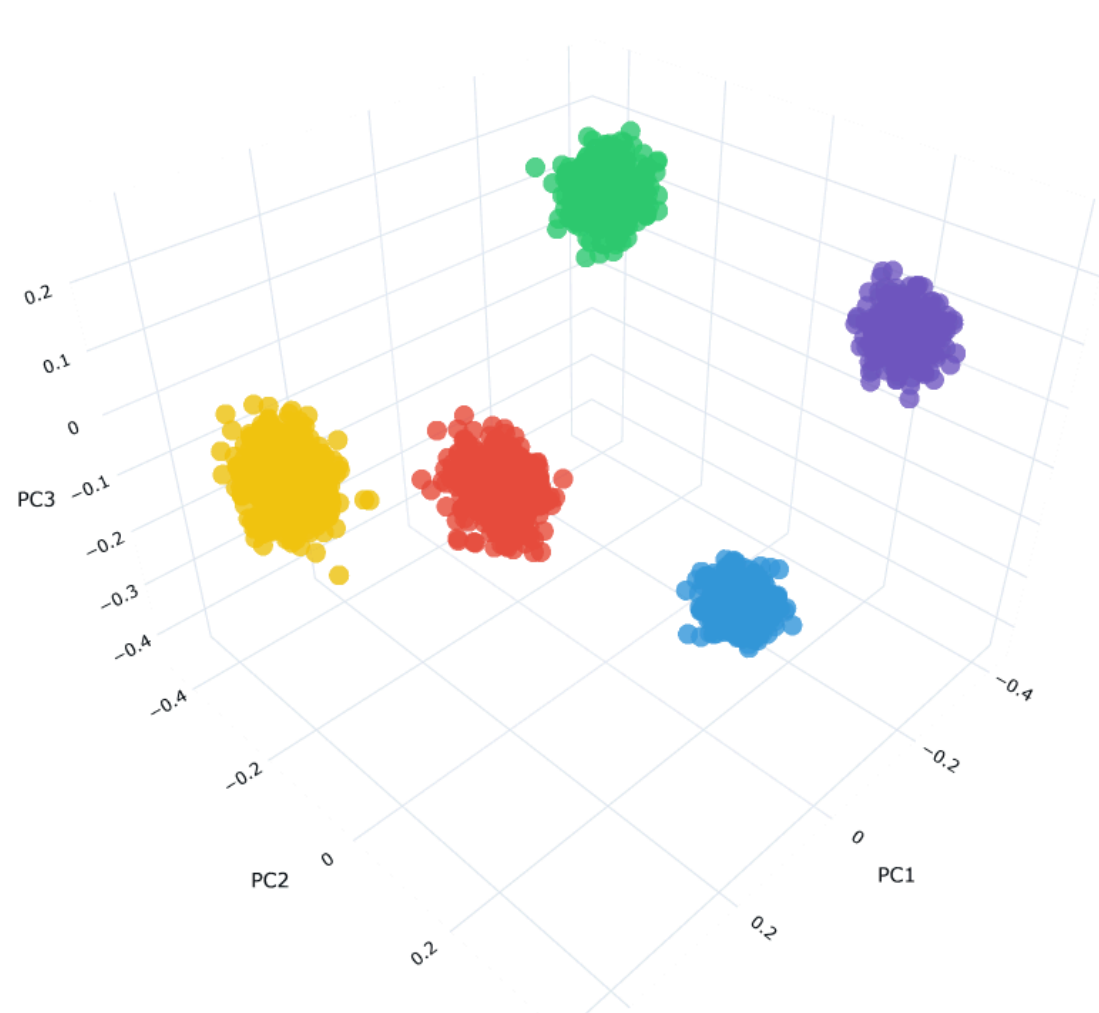

**Figure 23.** 3D PCA projection for multimodal Gaussian data ($k$ = 5).

### 3.5 Clustering Empirical Data

For the empirical dataset, classical K-means clustering indicated $k$ = 2 as the best solution according to the silhouette criterion (Figure 24). The average silhouette value was approximately 0.31, suggesting a modest but interpretable separation between clusters. The solution was also highly stable across runs, with a mean ARI of 0.999 and a standard deviation of 0.001, indicating that the same partition was almost always recovered despite random initialization.

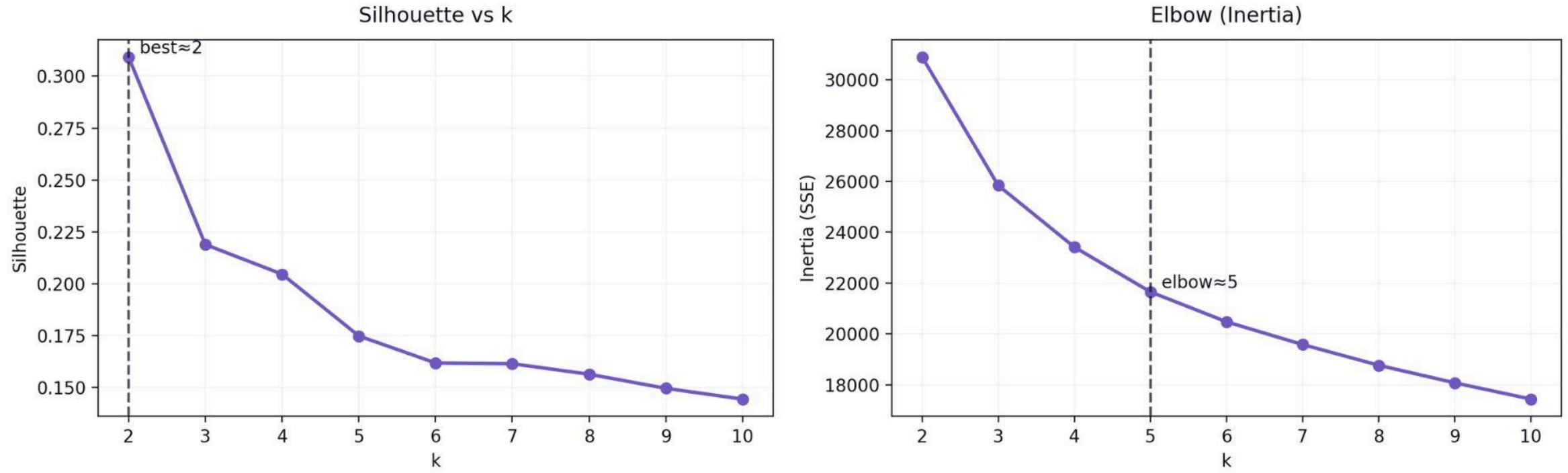

**Figure 24.** Silhouette and SSE vs k for empirical data.

The cluster profiles showed a clear and consistent difference in overall standardized scores across all selected psychological measures (Figure 25). Cluster 0 presented systematically lower scores across STARS, R-TAS, LSAS-SR, CAS, IUS-SF, and BFNE, whereas cluster 1 presented systematically higher scores across the same variables. The two clusters were nearly equal in size, with 4,237 participants in cluster 0 (50.6%) and 4,123 participants in cluster 1 (49.4%). This pattern suggests that the empirical solution reflects a broad low-symptom versus high-symptom profile, rather than highly specific subtypes or psychological phenomena.

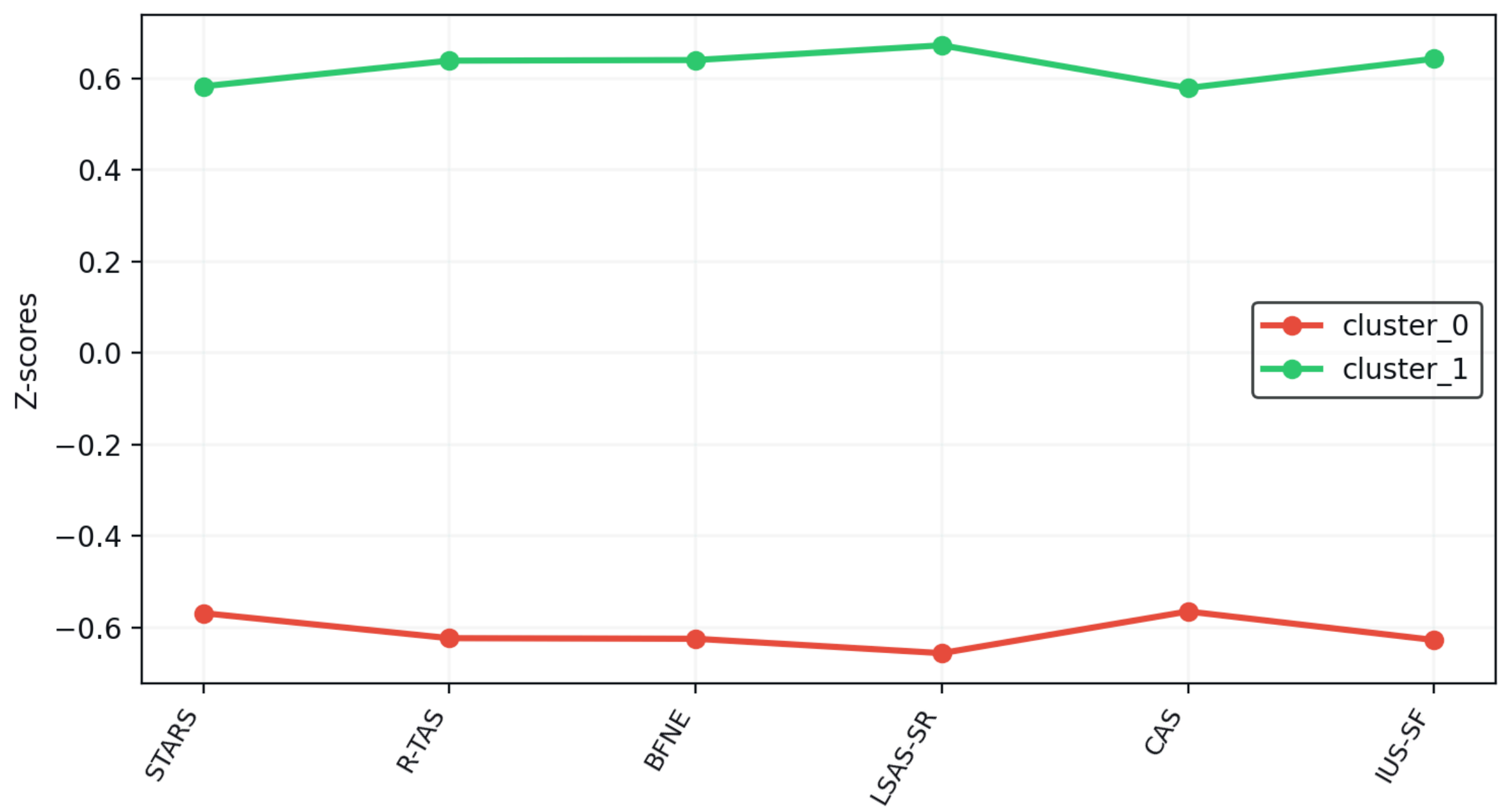

**Figure 25.** Profile-plot for empirical data ($k = 2$).

For the empirical dataset, spherical K-means clustering also selected k = 2 as the best solution (Figure 26). The cosine silhouette value was approximately 0.51, indicating a clearer angular separation than that observed with classical K-means. The solution was extremely stable across runs, with a mean ARI of 1.00 and a standard deviation of 0.000.

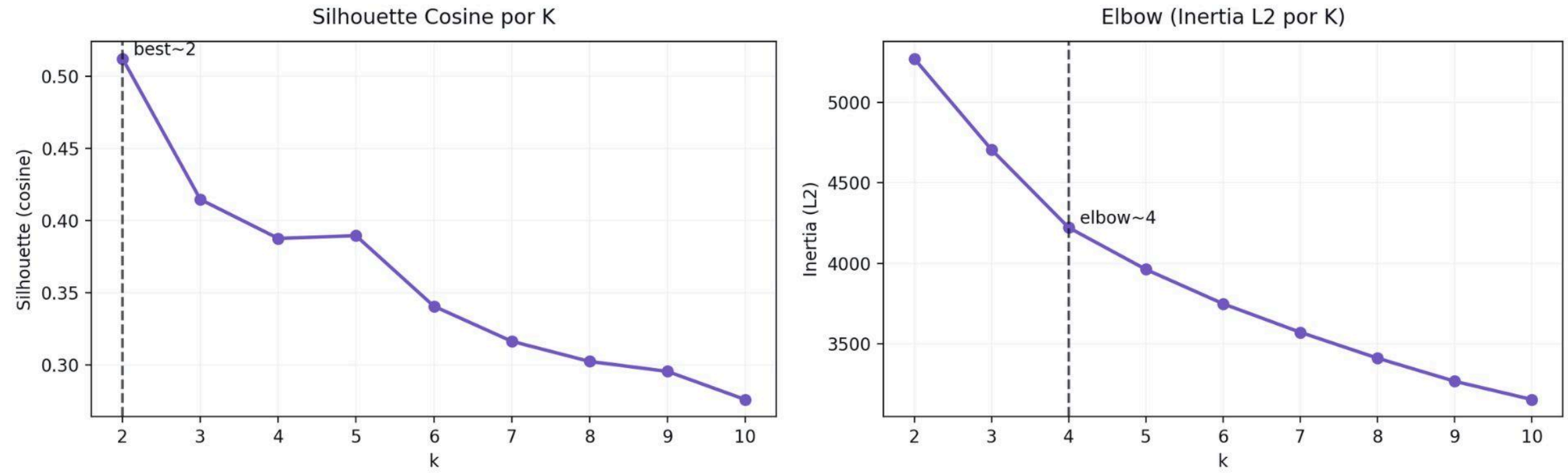

**Figure 26.** Silhouette (cosine) and SSE (L2) vs $k$ for empirical data.

Again, these clusters should not be interpreted as natural psychological profiles in a strict taxonomic sense, as revealed by the 2D PCA projection (Figure 27). Rather, they represent geometric boundaries drawn within a latent continuous psychological space. For this reason, the resulting groups are partly artificial: they are useful for summarizing patterns in the data, but they do not necessarily correspond to discrete, naturally occurring subpopulations.

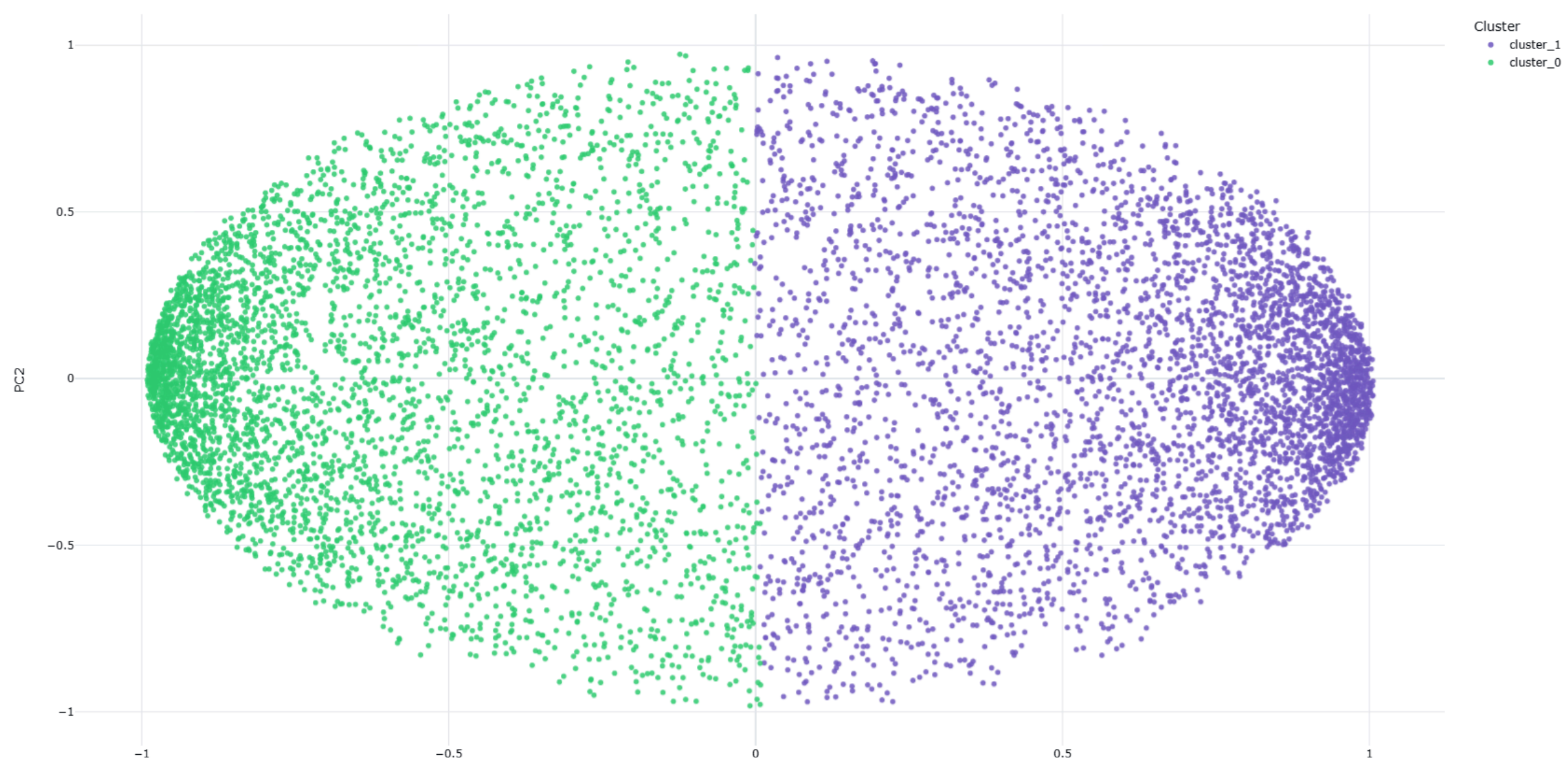

**Figure 27.** 2D PCA projection for empirical data ($k = 2$).

This interpretation is reinforced by the silhouette results and PCA projection for $k = 2, 3, 4$. Although $k = 2$ was selected as the best solution, additional partitions still produced reasonably comparable silhouette values, suggesting that the psychological space can be subdivided in several geometrically plausible ways (Figure 28). This is exactly what was observed in the simulation with unimodal correlated Gaussian dataset.

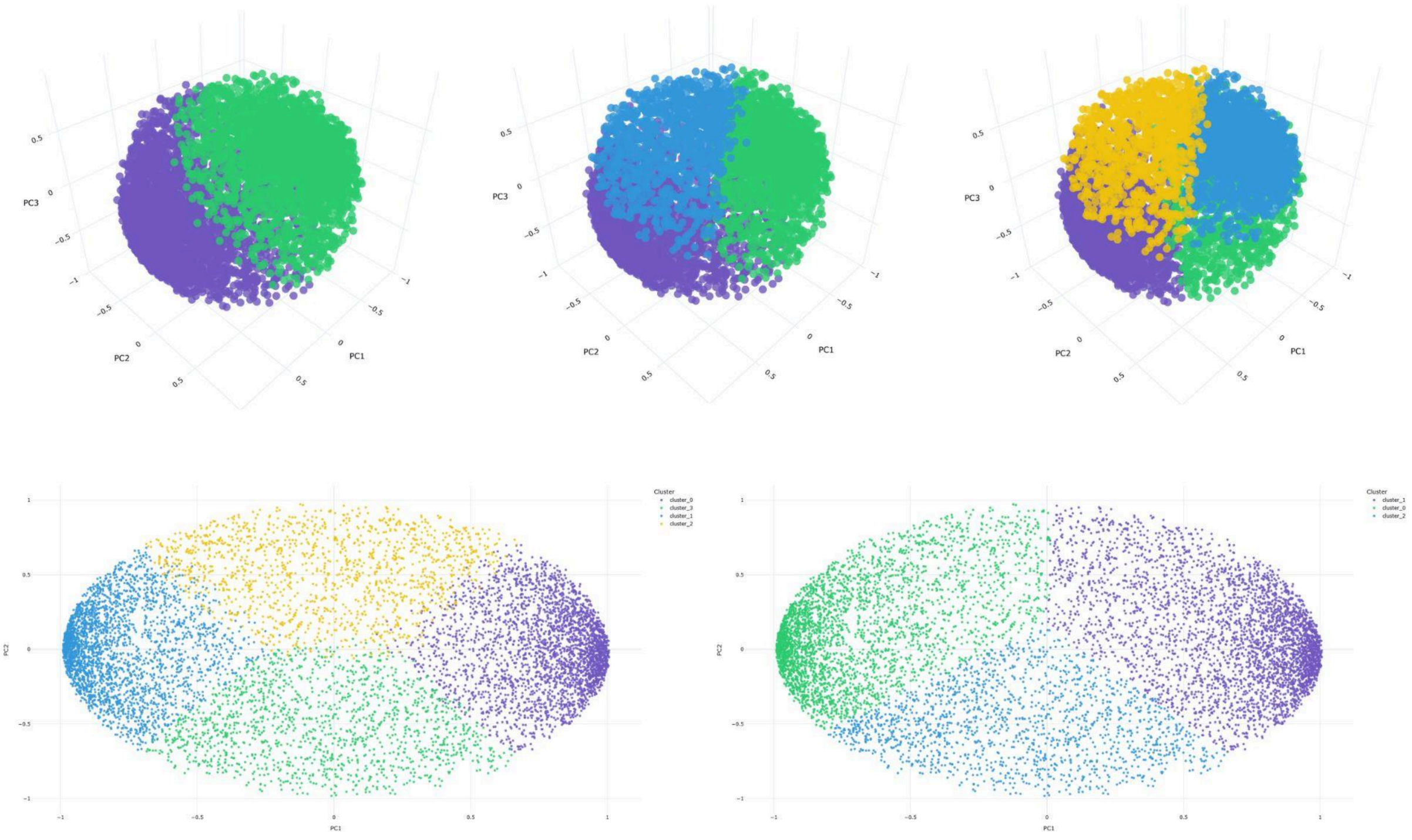

**Figure 28.** PCA projections for empirical data (k = 2, 3, 4).

### 3.6 Clustering Cytometer-like Data

For the simulated flow-cytometry-like dataset, classical K-means selected $k = 2$ as the best solution according to the silhouette criterion (Figure 29). Unlike the random and unimodal Gaussian conditions, this result was supported by a relatively high silhouette value, reaching approximately 0.52, which indicates a much stronger degree of separation between observations and their assigned clusters. The clustering solution was also perfectly stable across repeated runs (ARI = 1.000; SD = 0.000), showing that the same partition was consistently recovered despite different random initializations.

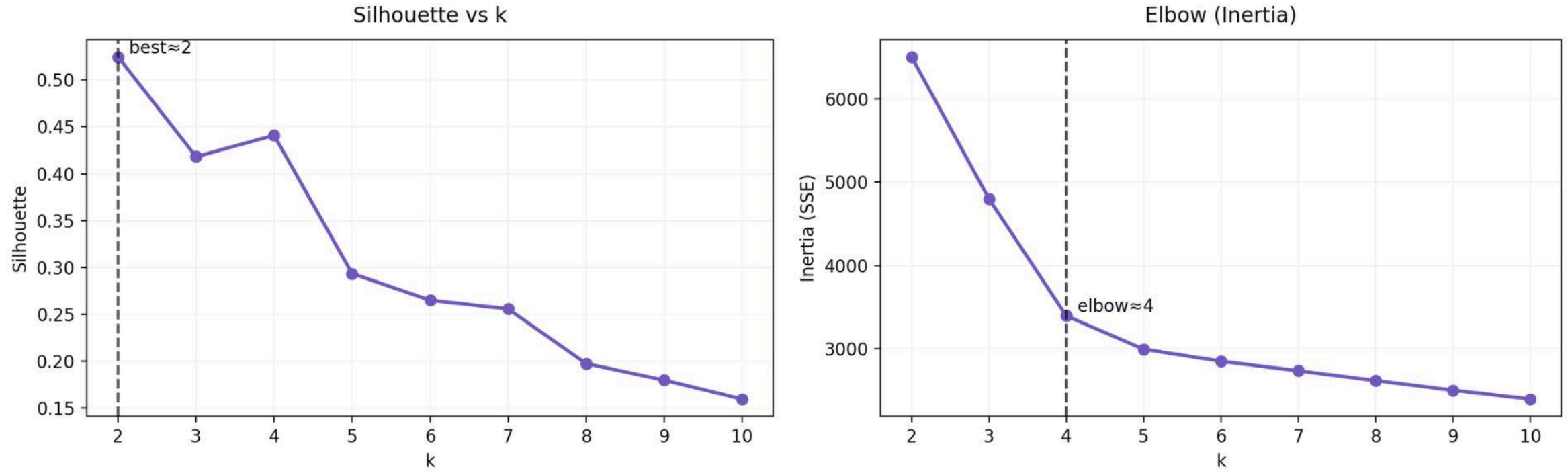

**Figure 29.** Silhouette and SSE vs k for cytometer-like data.

However, the silhouette and inertia criteria did not fully converge: while the silhouette curve favored $k = 2$, the inertia curve suggested a more plausible elbow around $k = 4$ (Figure 29). This discrepancy indicates that the strongest global separation in the dataset occurred at a broad two-cluster level (Figure 30), whereas additional internal structure was still present and became more visible through the reduction in SSE.

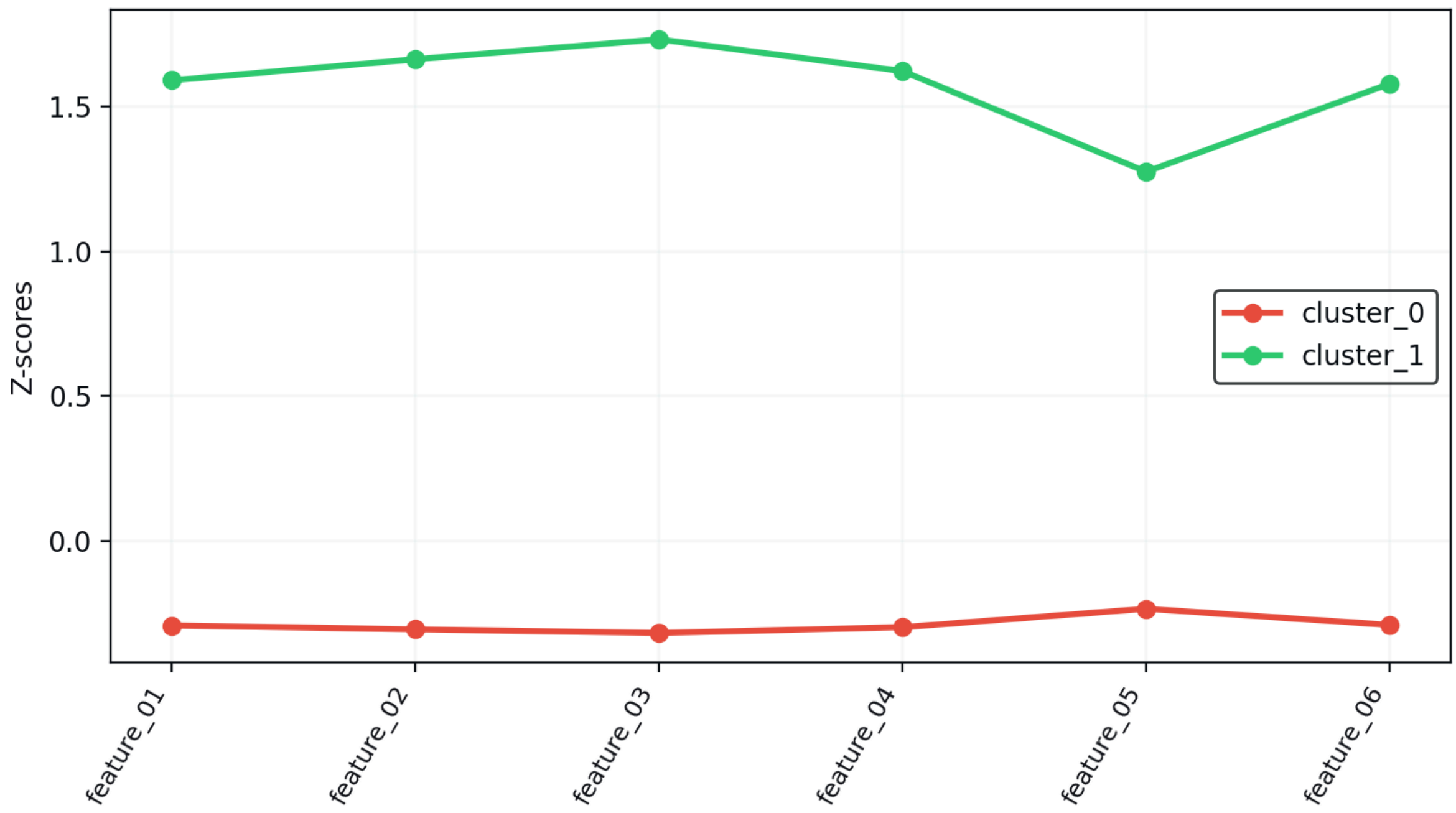

**Figure 30.** Profile-plot for cytometer-like data ($k = 2$).

When spherical K-means was applied to the same simulated cytometry data, the cosine silhouette criterion selected $k$ = 4 as the best solution (Figure 31). This solution was highly stable, with a mean ARI of 0.996 and a standard deviation of 0.003. The inertia curve also supported the presence of meaningful structure, although it did not point to exactly the same value of $k$. In this sense, the spherical K-means result suggests that the dataset contains not only a dominant global separation, but also finer internal subdivisions that can be recovered depending on the distance geometry and the criterion used to select $k$.

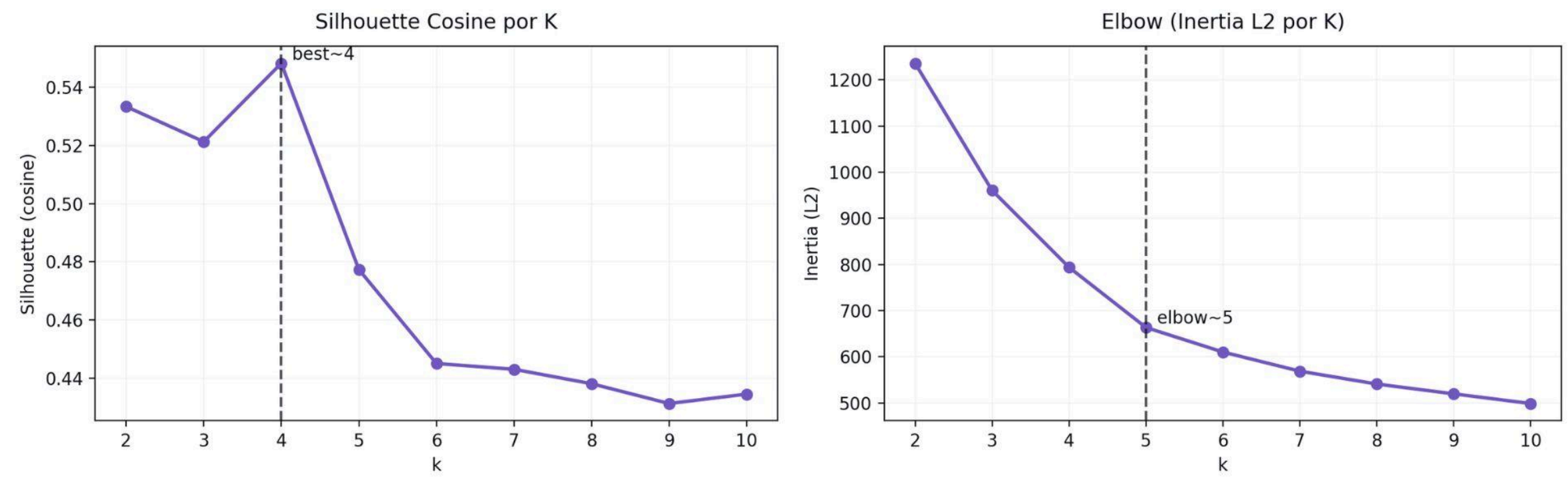

**Figure 31.** Silhouette (cosine) and SSE (L2) vs k for cytometer-like data.

In the PCA projection, spherical K-means revealed clearly distinguishable clusters in the two-dimensional space (Figure 32). Unlike the random-data condition and the Gaussian simulations, whether correlated or uncorrelated, the projected groups did not appear as arbitrary subdivisions of a continuous cloud. Instead, the clusters occupied relatively distinct regions of the PCA space, with only limited overlap between some groups.

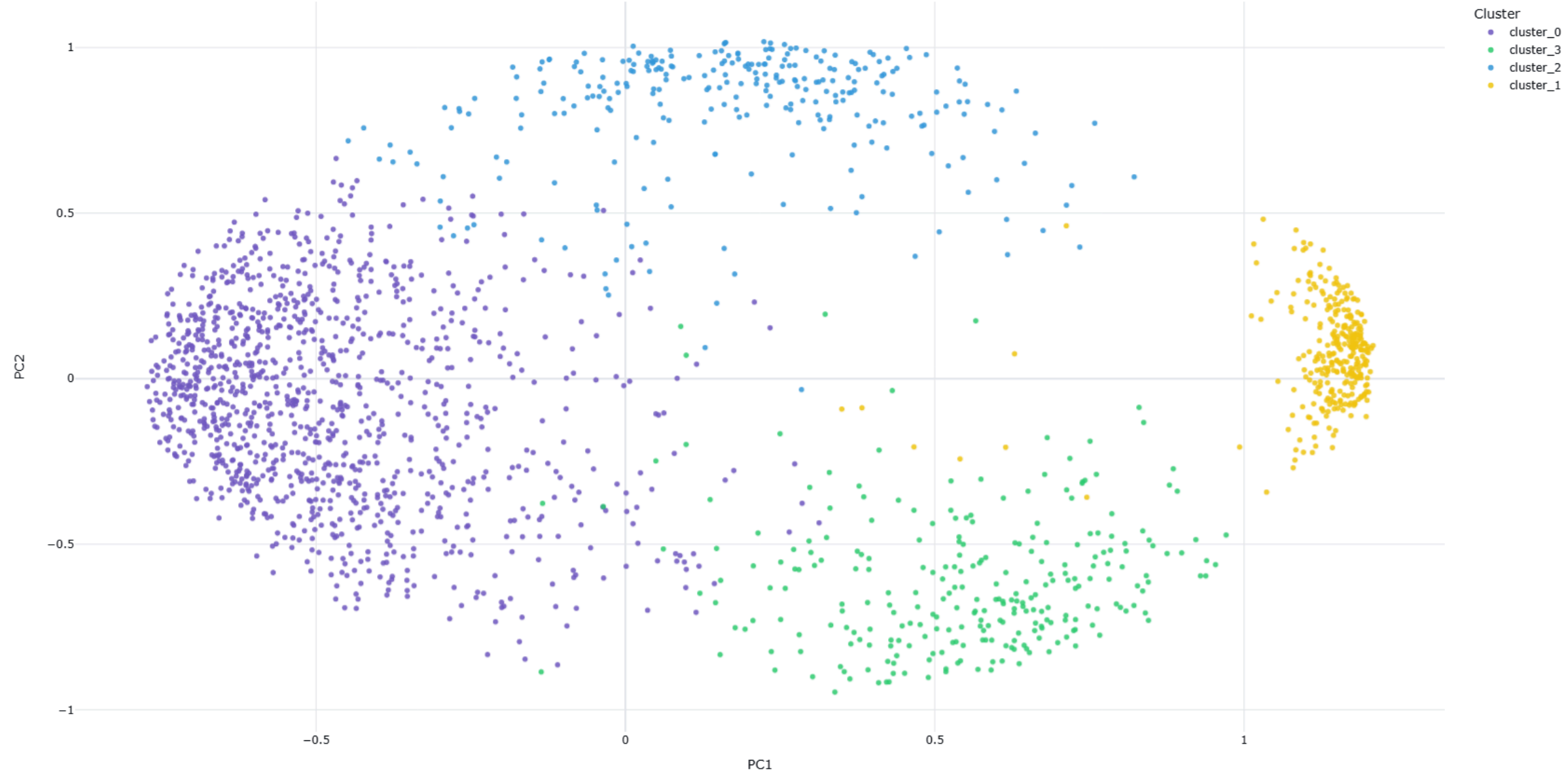

**Figure 32.** 2D PCA projection for cytometer-like data ($k$ = 4).

Even so, this example also shows that the choice of $k$ remains partly interpretive even under favorable simulation conditions. Although the dataset was generated to imitate five cell-like populations in a flow-cytometry context, the statistical criteria did not fully recover all five intended groups as the single optimal solution. Instead, the classical K-means solution emphasized a broader two-group separation, whereas spherical K-means captured a more fine-grained four-cluster structure. This discrepancy illustrates that the "best" value of $k$ depends not only on the presence of underlying structure, but also on the geometric criterion used by the algorithm and on the level of resolution considered meaningful by the researcher The intended five-cluster structure becomes clearer in the 3D PCA projection (Figure 33).

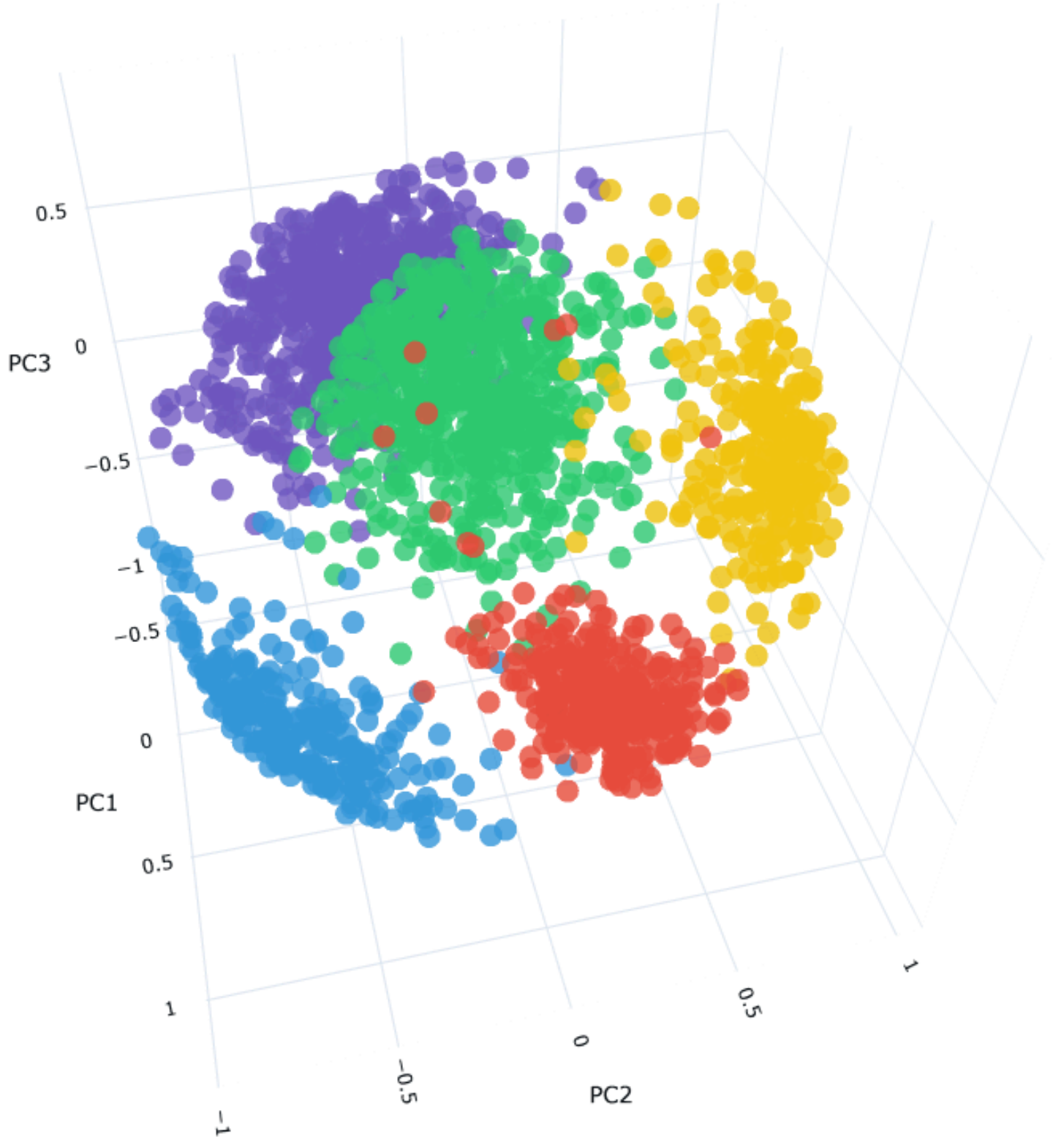

**Figure 33.** 3D PCA projection for cytometer-like data ($k$ =5).

In this case, the failure to recover exactly five clusters should not be interpreted as an error of the method. Two simulated cell populations were intentionally placed close to one another, making them statistically and geometrically similar. As a result, a generic clustering criterion could reasonably merge them, even though they were defined as separate populations in the data-generating process.

## 4 Discussions

The present analyses show that K-means can produce visually coherent and statistically stable clusters across very different data-generating conditions, but that the meaning of those clusters depends strongly on the structure of the data. In the random and unimodal Gaussian simulations, the algorithm still returned partitions that could be described as profiles, even though no true subgroup structure was present. This is a central limitation of K-means in psychological research: the method will always divide the feature space when asked to do so. Stability, therefore, is not equivalent to validity.

A solution can be also highly reproducible across random initializations and still represent only a geometric segmentation of continuous variation, as portrayed in the unimodal correlated Gaussian data clustering. When variables were correlated, the data formed an elongated latent structure rather than compact, naturally separated groups. These results illustrate how K-means can transform continuous psychometric variation into artificial categories. Because many psychological variables reflect the aggregate influence of multiple partially independent factors, approximately continuous Gaussian latent spaces may be statistically expected, consistent with the logic of the Central Limit Theorem[1]. A cluster solution may look like a "low profile" and a "high profile," but this does not necessarily mean that the sample contains two natural psychological phenotypes. It may simply mean that the algorithm has drawn a boundary through a continuous psychological gradient.

The multimodal Gaussian simulation provided the opposite case. Here, K-means was not misleading: the data were generated from two distinct Gaussian components, and classical K-means recovered this structure almost perfectly. This demonstrates that the critique is not that K-means is inherently invalid. Rather, K-means is most interpretable when the data resemble its implicit assumptions: compact groups, relatively clear separation, and meaningful differences in Euclidean position [Steinley, 2006]. In this context, clustering can also be useful because multimodality can distort standard statistical summaries. Moreover, strong multimodality violates the usual expectation of approximately unimodal and normally distributed variables or residuals in many parametric analyses.

The comparison between classical and spherical K-means further shows that different clustering algorithms answer different geometric questions. Classical K-means emphasizes Euclidean distance and therefore tends to separate observations according to overall level or position in the feature space. Spherical K-means, by contrast, emphasizes cosine similarity and is more sensitive to direction or profile shape [Hornik et al., 2012; Dhillon and Modha, 2001]. This can be useful when the substantive question concerns relative configurations across variables rather than absolute severity. However, the simulations also show that spherical K-means can subdivide the same latent space into additional angular regions, producing stable solutions that are not necessarily more natural. A higher or comparable silhouette value alone should therefore not be treated as automatic evidence of better psychological validity, even when the result is perfectly reproducible across initializations (ARI = 1.000; SD = 0.000).

The simulated cytometry dataset helps clarify this distinction. In a flow-cytometry-like context, clustering is often applied to data where biologically meaningful populations are expected. In that scenario, K-means behaves much like a trained human observer would when visually inspecting a cytometry plot: it identifies dense regions and separates known cell-like populations. This is important because, despite the recent popularization of Artificial

[1] The Central Limit Theorem states that when a variable reflects the combined influence of many partially independent factors, the resulting distribution tends to approximate a Gaussian form [Kwak & Kim, 2017]. In psychological measurement, where traits and symptoms often emerge from multiple biological, cognitive, social, and environmental influences, approximately continuous and correlated latent spaces may therefore be statistically expected.

Intelligence (AI) through Large Language Models (LLMs) such as ChatGPT, many machine-learning methods still approximate basic forms of pattern recognition that humans routinely perform in applied settings. Historically, several computational methods, including neural networks, were inspired by attempts to imitate aspects of human cognition [Hassabis et al., 2017]. K-means is much simpler than a neural network, but in a laboratory-like cytometry setting it captures a similar intuition: nearby events with similar multivariate measurements are grouped together.

At the same time, the cytometry simulation also shows that even under ideal conditions the choice of $k$ remains partly subjective. The dataset was generated to imitate five cell types, but the silhouette criterion did not necessarily recover all five intended populations. Instead, the best solution merged two highly similar cell types, which was intentional in the simulation. This reflects a realistic biomedical situation: some cell populations are known to be biologically distinct but occupy neighboring regions of the measurement space [Lötsch & Ultsch, 2024]. A generic clustering index may reasonably merge them, whereas a trained biomedical analyst may separate them because of prior biological knowledge. Thus, the "optimal" $k$ is not only a statistical question. It also depends on the level of resolution required by the scientific context.

The empirical SMARVUS results should be interpreted with this caution in mind. Both classical and spherical K-means produced highly stable two-cluster solutions, with balanced groups and coherent low-versus-high profiles across psychological measures. This suggests that the algorithms identified a robust organization of the sample. However, the profile structure resembles a broad severity gradient more than a set of sharply bounded psychological subtypes. The clusters are therefore better understood as lines drawn through a latent psychological space than as natural categories. Dividing the sample further would likely produce additional interpretable profiles, and some of these solutions could still show acceptable or comparable silhouette values.

This is exactly what the correlated Gaussian simulation demonstrated: continuous structures can be partitioned repeatedly into stable, visually meaningful, but ultimately artificial regions. A very similar pattern also emerged in our analyses using the SMARVUS educational-context dataset, where K-means consistently produced high-versus-low profile solutions resembling those commonly reported in the psychopathology literature. For example, Croci et al. [2026] reported a two-cluster solution for NSSI profiles composed of approximately equally sized groups, essentially splitting the sample into high and low symptom configurations, with a silhouette coefficient around 0.31; remarkably similar to the solution obtained in our analyses with the SMARVUS dataset.

## 5 Conclusion

The present study shows that K-means can generate stable and visually interpretable partitions even when the underlying data do not contain true latent groups. Across simulations, the algorithm performed as expected when the data were genuinely multimodal or clearly separated, as in the multimodal Gaussian and simulated cytometry conditions.

However, when applied to random, unimodal, or correlated Gaussian data, K-means still produced apparently coherent clusters by imposing artificial boundaries on continuous multivariate spaces. This demonstrates that stability, interpretability, and even acceptable silhouette values are not sufficient evidence that clusters correspond to natural psychological profiles.

In the empirical SMARVUS data, both classical and spherical K-means identified highly stable two-cluster solutions, with balanced group sizes and coherent low-versus-high psychological profiles. Nevertheless, these results are better interpreted as geometric stratifications of a latent psychological continuum rather than as evidence for discrete subtypes. The simulations showed that similar partitions can emerge from continuous correlated data, meaning that empirical cluster solutions must be interpreted cautiously, especially in psychometric datasets composed of linear associations.

K-means remains useful as an exploratory and descriptive tool. It can summarize complex data, reveal broad regions of multivariate space, and support hypothesis generation. However, it should not be treated as a method for validating latent psychological categories on its own. Researchers using clustering in psychological and social-determinant data should report cluster sizes, model-selection indices across multiple values of k, and visualizations of the data structure, while avoiding circular post-clustering inference. Ultimately, K-means clustering within linear data should be understood less as a discovery machine for natural subtypes and more as a way of drawing interpretable, but artificial, lines through complex psychological space.